\documentclass[structabstract]{aa}  
\pdfoutput=1 
\usepackage{natbib, graphicx}
\usepackage{amsmath}
\usepackage[flushleft]{threeparttable}
\usepackage{color}
\usepackage{txfonts}
\usepackage{multirow}
\usepackage{booktabs}
\usepackage{pdflscape}

\begin{document}

   \title{Spectroscopic parameters for solar-type stars with moderate/high rotation.}
\subtitle{New parameters for 10 planet-hosts\thanks{Based on 
observations collected at the La Silla Observatory, ESO (Chile) with FEROS spectrograph at the 2.2\,m 
telescope (ESO runs ID 089.C-0444(A), 088.C-0892(A)) and HARPS spectrograph at the 3.6\,m telescope (ESO runs ID 072.C-0488(E), 
079.C-0127(A)), at the Observatoire de Haute-Provence (OHP, CNRS/OAMP), France, with SOPHIE spectrograph at the 1.93\,m telescope 
and at the the Observatoire Midi-Pyr\'{e}n\'{e}es (CNRS), France, with NARVAL spectrograph at the 2\,m Bernard Lyot Telescope 
(Run ID L131N1).}}
   \author{M. Tsantaki\inst{1,2} 
          \and    
          S. G. Sousa\inst{1,2}          
         \and    
          N. C. Santos\inst{1,2}
	  \and 
	  M. Montalto\inst{1}
	  \and 
	  E. Delgado-Mena\inst{1}
	  \and
	  A. Mortier\inst{1}
	  \and 
	  V. Adibekyan\inst{1}
	  \and 
	  G. Israelian\inst{3}}
   \institute{Centro de Astrof\'{i}sica, Universidade do Porto, Rua das Estrelas, 4150-762 Porto, Portugal \\
              \email{Maria.Tsantaki@astro.up.pt} 
     \and
        Departamento de F\'{i}sica e Astronomia, Faculdade de Ci\^encias, Universidade do Porto, Rua do Campo Alegre, 
4169-007 Porto, Portugal
   \and
         Instituto de Astrof\'{i}sica de Canarias, E-38200 La Laguna, Tenerife, Spain 
       }

   \date{Received XXXX; accepted XXXX}
   \authorrunning{M. Tsantaki et al.}
   %\titlerunning{bla}
   \abstract
{Planetary studies demand precise and accurate stellar parameters as input to infer the planetary properties. 
Different methods often provide different results that could lead to biases in the planetary parameters.}
{In this work, we present a refinement of the spectral synthesis technique designed to treat better more rapidly 
rotating FGK stars. This method is used to derive precise stellar parameters, namely effective 
temperature, surface gravity, metallicitity and rotational velocity. This procedure is tested for samples of low and moderate/fast 
rotating FGK stars.}
{The spectroscopic analysis is based on the spectral synthesis package Spectroscopy Made Easy (SME), assuming Kurucz model 
atmospheres in LTE. The line list where the synthesis is conducted, is comprised of iron lines and the atomic data are derived 
after solar calibration.}
{The comparison of our stellar parameters shows good agreement with literature values, both for low and for higher rotating stars. 
In addition, our results are on the same scale with the parameters derived from the iron ionization and excitation method presented in 
our previous works. 
We present new atmospheric parameters for 10 transiting planet-hosts as an update to the SWEET-Cat catalogue. 
We also re-analyze their transit light curves to derive new updated planetary properties.}
{}

\maketitle

\section{Introduction}
\label{intro}

Since the first discoveries of the extrasolar planets, it became clear that the derivation of their fundamental properties was directly 
linked to the properties of their host stars.
%Since the discovery of the first extrasolar planet orbiting a main sequence star \citep{Mayor95}, more than 1\,700 planets\footnote{according to the online catalogue www.exoplanet.eu} have been confirmed, most of them around FGK stars. 
Until recently the discovery of extrasolar planets was substantially fed by the radial velocity (RV) technique. In the last years, several 
space missions such as CoRoT \citep{baglin2006} and \textit{Kepler} \citep{borucki2010}, 
as well as ground based surveys like WASP \citep{pollacco2006}) and HAT-P \citep{bakos2004} are successfully using the transit technique. 
The large number of planets discovered today\footnote{More than 1800 planets have been discovered up-to-date according to the 
online catalogue: www.exoplanet.eu}, allows the study of correlations in the properties of planets and their parent stars, providing 
strong observational constrains on the theories of planet formation and evolution \citep[and references therein]{mordasini12}.

%These transiting planets show a large diversity in their properties and along with the RV planets, provide 
%strong observational constrains on the theories of planet formation and evolution \citep[and references therein]{mordasini12}.

To understand the physical processes involved in the formation and evolution of planetary systems, precise measurements of the 
fundamental properties of the exoplanets and their hosts are required. From the analysis of the light curve of a transiting 
planet, the planetary radius determination is always dependent on the stellar radius (R$_{p}$ $\propto$ R$_{\star}$). 
Moreover, the mass of the planet, or the minimum mass in case the inclination of the orbit is not known, is calculated from 
the RV curve only if the mass of the star is known (M$_{p}$ $\propto$ M$_{\star}^{2/3}$). The fundamental stellar 
parameters of mass and radius, on the other hand, depend on observationally determined parameters such as effective temperature 
($T_{\mathrm{eff}}$), surface gravity ($\log g$), and metallicitity ($[Fe/H]$, where iron is usually used as a proxy). 
The latter fundamental parameters are used to deduce stellar mass and radius either from calibrations 
\citep{torres10_mass, santos13} or stellar evolution models \citep[e.g.][]{girardi2002}.

It is therefore, imperative to derive precise and accurate stellar parameters to avoid the propagation of errors in the 
planetary properties. For instance, \cite{torres12} have shown that unconstrained parameter determinations derived from 
spectral synthesis techniques introduce considerable systematic errors in the planetary mass and radius. In particular, 
residual biases of the stellar radius may explain part of the anomalously inflated radii that has been observed for some 
Jovian planets such as in the cases of HD\,209458\,b \citep{burrows00} and WASP-12\,b \citep{hebb09}.

There are several teams applying different analysis techniques (e.g. photometric, spectroscopic, interferometric), atomic data, model 
atmospheres, etc., and their results often yield significant differences \citep[e.g.][]{torres08, bruntt2012, molenda13}. These 
systematic errors are difficult to assess and are usually the main error contributors within a study. 
Such problems can be mitigated by a uniform analysis that will yield the precision needed. Apart from minimizing the 
errors of the stellar/planet parameters, uniformity can enhance the statistical significance of correlations between the presence of 
planets and the properties of their hosts. For example, an overestimation in the stellar radius has been reported in some 
samples of \textit{Kepler} Objects of Interest \citep{verner2011,everett2013} which in turn leads to overestimated planetary 
radius. In this case, planets are perhaps misclassified in the size range likely for rocky Earth-like bodies, affecting the 
planet occurrence rate of Earth-sized planets around solar-type stars.

The high quality stellar spectra obtained from RV planet search programs \citep[e.g.][]{sousa1, sousa3}, make 
spectroscopy a powerful tool for deriving the fundamental parameters in absence of more direct radius measurements 
(restricted only to limited stars with the interferometric technique or stars that belong to eclipsing binaries). A typical 
method of deriving stellar parameters for solar-type stars is based on the excitation and ionization equilibrium by measuring 
the equivalent widths (EW) of iron lines (hereafter EW method). This method has successfully been applied to RV targets that 
are restricted to low rotational velocities ($\upsilon\sin i$) to increase the precision of the RV technique \citep{Bouchy01}. 
High rotational velocities also limit the precision of the EW method. Spectral lines are broadened by rotation and therefore 
neighboring lines become blended, often unable to resolve. Even though the EW is preserved, its correct measurement is not yet 
possible.

On the other hand, the transit planet-hosts have a wider dispersion in rotation rates when comparing to the slow rotating FGK hosts 
observed with the RV technique. For moderate/high rotating stars, which may be the case of the transit targets, spectral synthesis is 
required for the parameter determination. This technique yields stellar parameters by fitting the observed spectrum with a 
synthetic one \citep[e.g.][]{valenti05, malavolta2014} or with a library of pre-computed synthetic spectra \citep[e.g.][]{recio06}. 

In this paper, we propose a refined approach based on the spectral synthesis technique to derive stellar parameters 
for low-rotating stars (Sect.~\ref{method}), yielding results on the same scale with the homogeneous analysis of our previous 
works (Sect.~\ref{results1}). Our method is tested for a sample of moderate/high FGK rotators (Sect.~\ref{rotation}) and also 
is applied to a number of planet-hosts providing new stellar parameters. Their planetary properties are also revised 
(Sect.~\ref{starsplanets}). 

\section{Spectroscopic analysis}\label{method}

\begin{figure}
  \centering
  \includegraphics[width=1.0\linewidth]{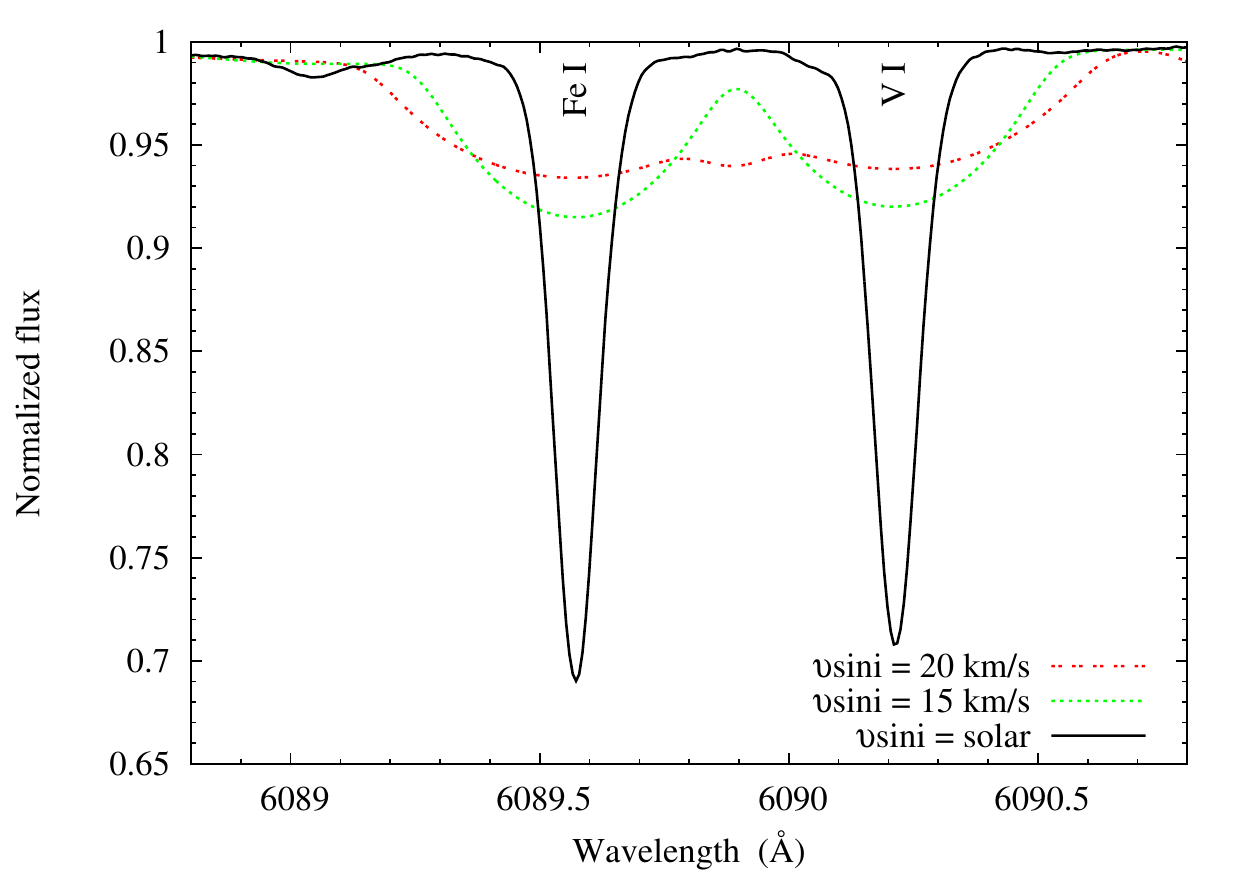}
  \caption{\footnotesize{Solar absorption lines (solid line), broadened by $\upsilon\sin i$ 15\,km/s (green) and 20\,km/s (red). 
Blending at these rates due to rotation makes the accurate measurement of the EW very difficult.}}
  \label{fig:1}
  \end{figure}

Due to severe blending, measuring the EW of stars with high rotational velocity is very difficult, if not impossible (e.g. see 
Fig.~\ref{fig:1}). 
%An approximate limit of the rotational velocity where the EW method provides reliable results, is up to $\upsilon\sin i$ $\sim$12-15\,km/s (depending on the choice of lines and spectral type). 
In this paper, we are focusing on deriving precise and accurate parameters for stars with higher $\upsilon\sin i$ using the 
spectral synthesis technique. 

%----------------------------------------- FIGURE ---------------------------------------------------
%--------------------------------------------------------------------------------------------------

\subsection{Line list}\label{linelist}

For an accurate spectral synthesis, atomic and molecular data of all lines in the wavelength intervals where the synthesis is 
conducted must be as accurate as possible. The choice of intervals for our analysis is based on the line list of iron lines as 
described in \cite{tsantaki13}. This list is comprised of weak, isolated iron lines, specifically chosen from the extended line list of 
\cite{sousa1} to exclude blended lines that are commonly found in K-type stars. Effective temperatures derived with this line 
list are in agreement with the InfraRed-Flux Method (IRFM) for the whole temperature regime of FGK dwarfs.

The spectral window around each iron line is set wide enough to include broadened lines of 
$\upsilon\sin i$\,$\sim$\,50\,km/s. Following the Doppler law, such a rotational velocity causes a broadening of 
$\pm$\,1\,\AA{}, around a line in the middle of the optical wavelength range ($\sim$ 5500\,\AA{}).

%\textbf{We increased the wavelength interval to $\pm$\,5\,\AA{} to include enough continuum points in each of the 
%defined intervals around the $\ion{Fe}{i}$ and $\ion{Fe}{ii}$ lines. However, it is computational expensive. }

The original line list contains 137 $\ion{Fe}{i}$ and $\ion{Fe}{ii}$ lines where we set intervals of 2\,\AA{} around 
them. The atomic data for these intervals were obtained from the Vienna Atomic Line 
Database\footnote{http://vald.inasan.ru/$\sim$vald3/php/vald.php} \citep{piskunov95, kupka99}. We extracted atomic data for all 
the expected transitions for a star with solar atmospheric parameters for our wavelength intervals. We also included lines 
predicted for a K-type star with $T_{\mathrm{eff}}$\,=\,4400\,K. The two line lists that correspond to atomic transitions 
for the two different spectral types were merged into one after removing duplicates. Molecular data of the most abundant 
molecules in solar-type stars (C$_{2}$, CN, OH, and MgH) were also obtained from VALD using the same requests as for the atomic 
data. 

From the above intervals we selected the optimal ones according to the following procedure. From the first analyses, 
we noticed that K-type stars show the highest residuals 
between the observed and the best-fit synthetic spectrum compared to the F and G spectral types. The main reason is that the 
spectra of K-type stars include numerous lines but not all appear in our line list after the requested atomic data queries. 
%In addition, the observed with the synthetic spectrum in some intervals could not match properly because the continuum was not well defined. 
Therefore, we discarded lines in the bluer part (below 5000\,\AA{}) where lines are more crowded. Lines within overlapping 
intervals were merged into one.

In addition, we checked the behaviour of the remaining lines due to rotation by using the Sun as a reference star 
convolved with moderate rotation of 20\,km/s (see Sect.~\ref{rotation}). We excluded lines by eye where there was strong 
contamination by neighboring lines due to broadening and chopped the intervals were the contamination in the edges was weak. 
We also excluded lines that showed high residuals between the spectrum and the synthesized one for the solar parameters.

The initial choice of spectral windows was double the length i.e. 4\,\AA{} where the iron lines were placed in the 
center. For these intervals even though the best-fit parameters for the Sun (low rotation) were accurate, for the solar spectrum 
convolved with rotation (again of 20\,km/s), the parameters showed higher deviation ($T_{\mathrm{eff}}$ = 5728\,K, 
$\log g$ = 4.39\,dex, [Fe/H] = -0.03\,dex) compared to the standard solar values. For this reason we limited the length of the 
intervals to 2\,\AA{}.

Except for blending, another considerable problem that limits this procedure because of high rotation is the difficulty 
in distinguishing between the line and the continuum points as the lines become very shallow. In this case, the wings of the 
lines are miscalculated as continuum, leading to the biases in Sect~\ref{rotation}. 
%high rotation where the limitations of our method appears e.g. at 50\,km/s the lines are very shallow and close to the 
%continuum. To distinguish the continuum points from the line points, the difference has to be higher than 1\%.  }
%SME to define the lines. In that sense the intervals have just wide enough to include the line but continuum points as well. 
%(example) We want to avoid big intervals as they will make the computations longer. We also want to avoid short ones as 
%in cases with high rotation they will be considered as continuum.

%We extended the limits of each wavelength interval to $\pm$\,2\,\AA{} to 
%include enough continuum points in each of the defined intervals around the $\ion{Fe}{i}$ and $\ion{Fe}{ii}$ lines. 

Taking all the above into consideration, the final line list is comprised of 47 \ion{Fe}{i} and 4 \ion{Fe}{ii} lines 
into 42 wavelength intervals, summing in total of 537 lines of different species. The wavelength intervals and the atomic data 
of the iron lines are presented at Table~\ref{tab:line_data}.\footnote{The complete line list will be uploaded online in the 
SME format.}

%In some cases lines had to be added with new queries from VALD and intervals were chopped because neighboring lines were contaminating our intervals.

%------------------------------------------------------------------------------------------------------------------------------
\begin{table}
\centering
\caption{Spectral wavelength intervals and line data used for the spectroscopic analysis.}
\label{tab:line_data}
\begin{tabular}{cccccc}
\hline\hline
Intervals & $\lambda$ & Species & $\chi_{ex}$ &  $\log gf$ & $\varGamma_{6}$ \\
  \AA{}   &   \AA{}    &         &   (eV)      &            &  \\
\hline
5521.45 - 5523.45 & 5522.45 & \ion{Fe}{i} & 4.21 & -1.484 & -7.167 \\
5559.22 - 5561.11 & 5560.22 & \ion{Fe}{i} & 4.43 & -0.937 & -7.507 \\
5632.95 - 5634.95 & 5633.95 & \ion{Fe}{i} & 4.99 & -0.186 & -7.391 \\
5648.99 - 5652.47 & 5649.99 & \ion{Fe}{i} & 5.10 & -0.649 & -7.302 \\
 ...              & 5651.47 & \ion{Fe}{i} & 4.47 & -1.641 & -7.225 \\
 ...              & 5652.32 & \ion{Fe}{i} & 4.26 & -1.645 & -7.159 \\
5678.03 - 5680.97 & 5679.03 & \ion{Fe}{i} & 4.65 & -0.657 & -7.320 \\
 ...              & 5680.24 & \ion{Fe}{i} & 4.19 & -2.347 & -7.335 \\
5719.90 - 5721.90 & 5720.90 & \ion{Fe}{i} & 4.55 & -1.743 & -7.136 \\
5792.92 - 5794.92 & 5793.92 & \ion{Fe}{i} & 4.22 & -2.038 & -7.304 \\
5810.92 - 5812.92 & 5811.92 & \ion{Fe}{i} & 4.14 & -2.323 & -7.951 \\
5813.81 - 5815.45 & 5814.81 & \ion{Fe}{i} & 4.28 & -1.720 & -7.269 \\
 ...              & 5815.22 & \ion{Fe}{i} & 4.15 & -2.521 & -7.038 \\
5852.15 - 5854.15 & 5852.22 & \ion{Fe}{i} & 4.55 & -1.097 & -7.201 \\
 ...              & 5853.15 & \ion{Fe}{i} & 1.49 & -5.006 & -6.914 \\
5861.36 - 5863.36 & 5862.36 & \ion{Fe}{i} & 4.55 & -0.186 & -7.572 \\
5986.07 - 5988.07 & 5987.07 & \ion{Fe}{i} & 4.79 & -0.428 & -7.353 \\
6004.55 - 6006.55 & 6005.55 & \ion{Fe}{i} & 2.59 & -3.437 & -7.352 \\
6088.57 - 6090.57 & 6089.57 & \ion{Fe}{i} & 4.58 & -1.165 & -7.527 \\
6119.25 - 6121.25 & 6120.25 & \ion{Fe}{i} & 0.92 & -5.826 & -7.422 \\
6126.91 - 6128.78 & 6127.91 & \ion{Fe}{i} & 4.14 & -1.284 & -7.687 \\
6148.25 - 6150.25 & 6149.25 & \ion{Fe}{ii} & 3.89 & -2.786 & -7.478 \\
6150.62 - 6152.62 & 6151.62 & \ion{Fe}{i} & 2.18 & -3.188 & -7.729 \\
6156.73 - 6158.73 & 6157.73 & \ion{Fe}{i} & 4.08 & -1.097 & -7.691 \\
6172.65 - 6174.19 & 6173.34 & \ion{Fe}{i} & 2.22 & -2.775 & -7.829 \\
6225.74 - 6227.40 & 6226.74 & \ion{Fe}{i} & 3.88 & -2.021 & -7.423 \\
6231.65 - 6233.65 & 6232.65 & \ion{Fe}{i} & 3.65 & -1.161 & -7.552 \\
6237.00 - 6239.38 & 6238.39 & \ion{Fe}{ii} & 3.89 & -2.693 & -7.359 \\
6321.69 - 6323.69 & 6322.69 & \ion{Fe}{i} & 2.59 & -2.314 & -7.635 \\
6334.34 - 6336.34 & 6335.34 & \ion{Fe}{i} & 2.20 & -2.323 & -7.735 \\
6357.68 - 6359.68 & 6358.68 & \ion{Fe}{i} & 0.86 & -4.225 & -7.390 \\
6431.83 - 6433.05 & 6432.69 & \ion{Fe}{ii} & 2.89 & -3.650 & -7.391 \\
6455.39 - 6457.02 & 6456.39 & \ion{Fe}{ii} & 3.90 & -2.175 & -7.682 \\
6480.88 - 6482.88 & 6481.88 & \ion{Fe}{i} & 2.28 & -2.866 & -7.627 \\
6626.55 - 6628.55 & 6627.55 & \ion{Fe}{i} & 4.55 & -1.400 & -7.272 \\
6645.94 - 6647.50 & 6646.94 & \ion{Fe}{i} & 2.61 & -3.831 & -7.141 \\
6698.15 - 6700.15 & 6699.15 & \ion{Fe}{i} & 4.59 & -2.004 & -7.162 \\
6704.11 - 6706.11 & 6705.11 & \ion{Fe}{i} & 4.61 & -1.088 & 7.539 \\
6709.32 - 6711.32 & 6710.32 & \ion{Fe}{i} & 1.49 & -4.732 & -7.335 \\
6724.36 - 6727.67 & 6725.36 & \ion{Fe}{i} & 4.10 & -2.093 & -7.302 \\
 ...              & 6726.67 & \ion{Fe}{i} & 4.61 & -0.951 & -7.496 \\
6731.07 - 6732.50 & 6732.07 & \ion{Fe}{i} & 4.58 & -2.069 & -7.130 \\
6738.52 - 6740.52 & 6739.52 & \ion{Fe}{i} & 1.56 & -4.797 & -7.685 \\
6744.97 - 6746.97 & 6745.11 & \ion{Fe}{i} & 4.58 & -2.047 & -7.328 \\
 ...              & 6745.97 & \ion{Fe}{i} & 4.08 & -2.603 & -7.422 \\
6839.23 - 6840.84 & 6839.84 & \ion{Fe}{i} & 2.56 & -3.304 & -7.567 \\
6854.72 - 6856.72 & 6855.72 & \ion{Fe}{i} & 4.61 & -1.885 & -7.253 \\
6856.25 - 6859.15 & 6857.25 & \ion{Fe}{i} & 4.08 & -1.996 & -7.422 \\
 ...              & 6858.15 & \ion{Fe}{i} & 4.61 & -0.941 & -7.344 \\
6860.94 - 6862.94 & 6861.94 & \ion{Fe}{i} & 2.42 & -3.712 & -7.580 \\
 ...              & 6862.50 & \ion{Fe}{i} & 4.56 & -1.340 & -7.330 \\
\hline
\end{tabular}
\end{table}
%------------------------------------------------------------------------------------------------------------------

Atomic data are usually calculated from laboratory or semi-empirical estimates. In order to avoid uncertainties that may arise from 
such estimations, we determine astrophysical values for the basic atomic and molecular line data namely for the oscillator strengths 
($\log gf$) and the van der Waals damping parameters ($\varGamma_{6}$). We used the National Solar Observatory Atlas \citep{kurucz84} to 
improve the transition probabilities and the broadening parameters of our line list in an inverted analysis using the typical solar 
parameters fixed (as adopted by \cite{valenti05}: $T_{\mathrm{eff}}$\,=\,5770\,K, $\log g$\,=\,4.44\,dex, $[Fe/H]$\,=\,0.0\,dex, 
$\upsilon_{mic}$\,=\,0.87\,km/s, $\upsilon_{mac}$\,=\,3.57\,km/s, $\log_{\epsilon}(Fe)$ = 7.50\,dex). 

%\citep{holweger91}
\subsection{Initial conditions}

All minimization algorithms depend on the initial conditions. In order to make sure that the convergence is achieved for 
the global minimum, we set the initial conditions as close to the expected ones as possible. For temperature, we use the calibration 
of \cite{valenti05} as a function of \textit{B \textendash V} color. Surface gravities are calculated 
using \textit{Hipparcos} parallaxes \citep{hip}, \textit{V} magnitudes, bolometric corrections based on \cite{flower} and 
\cite{torres10}, and solar magnitudes from \citep{bessell}\footnote{Surface gravities from parallaxes are usually referred to as 
trigonometric in the literature.}. In cases the parallaxes are not available, we use the literature values. Masses are set to solar value. 

Microturbulence ($\upsilon_{mic}$) is used to remove possible trends in parameters due to model deficiencies. It has been shown that 
$\upsilon_{mic}$ correlates mainly with $T_{\mathrm{eff}}$ and $\log g$ for FGK stars \citep[e.g.][]{nissen, adib2, ramirez13}. We 
therefore, set $\upsilon_{mic}$ according to the correlation discussed in the work of \cite{tsantaki13} for a sample of FGK 
dwarfs. For the giant stars in our sample, we use the empirical calibration of \cite{mortier13} based on the results of 
\cite{hekker07}.

Macroturbulence ($\upsilon_{mac}$) is a broadening mechanism that also correlates with $T_{\mathrm{eff}}$ \citep[e.g.][]{saar1997}. 
We set $\upsilon_{mac}$  in our analysis following the relation of \cite{valenti05}. Initial metallicity ($[M/H]$) is set to solar and 
initial rotational velocity to 0.5\,km/s.

\subsection{Spectral synthesis}\label{synthesis}

The spectral synthesis package we use for this analysis is Spectroscopy Made Easy (SME), version 3.3 \citep{valenti96}. 
Modifications from the first version are described in \cite{valenti98} and \cite{valenti05}. The adopted model atmospheres are 
generated by the ATLAS9 program \citep{kurucz} and local thermodynamic equilibrium is assumed. SME includes the parameter 
optimization procedure based on the Levenberg-Marquardt algorithm to solve the nonlinear least-squares problem yielding the 
parameters that minimize the $\chi^{2}$. In our case, the free parameters are: $T_{\mathrm{eff}}$, $\log g$, $[M/H]$, and 
$\upsilon\sin i$. Metallicity in this work refers to the average abundance of all elements producing absorption in our 
spectral regions.  We can safely assume that $[M/H]$ approximately equals to $[Fe/H]$ for our sample of stars as the dominant lines in 
our regions are the iron ones. Additionally, these stars are not very metal-poor ($>$\,-0.58\,dex). The overall metallicity in metal-poor 
stars is enhanced by other elements (relative to iron) and in that case the previous assumption does not hold \citep[e.g.][]{adibekyan12}.

After a first iteration with the initial conditions described above, we use the output set of parameters to derive stellar 
masses using the Padova models\footnote{http://stev.oapd.inaf.it/cgi-bin/param} \citep{dasilva06}. Surface gravity is then re-derived 
with the obtained mass and temperature. The values of $\upsilon_{mic}$ and $\upsilon_{mac}$ are also updated by the new 
$T_{\mathrm{eff}}$ and $\log g$. The final results are obtained after a second iteration with the new initial values. Additional 
iterations were not required, as the results between the first and second iteration in all cases were very close (for instance the mean 
differences for the sample in Sect.\ref{rotation} are: $\Delta T_{\mathrm{eff}}$ =\,24\,K, $\Delta \log g$ =\,0.06\,dex, 
$\Delta$[Fe/H]\,=\,0.003\,dex and $\Delta \upsilon\sin i$ =\,0.18\,km/s).

\subsection{Internal error analysis}\label{error_analysis}

Estimation of the errors is a complex problem for this analysis. One approach is to calculate the errors from the covariance matrix of 
the best fit solution. Usually these errors are underestimated and do not include deviations depending on the specific choice of 
initial parameters nor the choice of the wavelength intervals. On the other hand, Monte Carlo approximations are computational 
expensive when we are dealing with more than a handful of stars. In this section we explore the contribution of different type of errors 
for reference stars of different spectral types. The errors of these stars will be representative of the errors of the whole group 
that each one belongs. 

We select 3 slow rotating stars of different spectral types: F (HD\,61421), G (Sun), and K (HD\,20868) as our references. Their stellar 
parameters are listed in Table~\ref{table8}. We convolve each of these stars with different rotational profiles (initial, 15, 25, 35, 45, 
and 55\,km/s) to quantitatively check the errors attributed to different $\upsilon\sin i$ (see also Sect~\ref{rotation}). 

Our aim is to calculate the errors from two different sources: 1) the initial conditions, and 2) the choice of wavelength 
intervals. Firstly, we check how the initial parameters affect the convergence to the correct ones. For each star we set 
different initial parameters by changing: $T_{\mathrm{eff}}$ $\pm$ 100 K, $\log g$ $\pm$ 0.20 dex, $[Fe/H]$ $\pm$ 0.10 dex, and 
$\upsilon\sin i$ $\pm$ 0.50 km/s. We calculate the parameters for the total 81 permutations of the above set of initial parameters. 
This approach is also presented in \cite{valenti05} for their solar analysis. 

The choice of wavelength intervals is also important for the precise determination of stellar parameters. 
The spectral window of different instruments varies and therefore not all wavelength intervals of a specific line list can be used 
for the parameter determination. Moreover, there are often other reasons for which discarding a wavelength interval would be wise, 
such as the presence of cosmic rays. In these cases, the errors which are attributed to the discarded wavelength intervals from a 
defined line list can give an estimation on the homogeneity of our parameters. 

We account for such errors by randomly excluding 10\% of our total number of intervals (that leaves us with 38 intervals). This percentage 
is approximately expected for the above cases. Stellar parameters are calculated for the shortened list of intervals and this procedure 
is repeated 100 times (each time discarding a random 10\%). The error of each free parameter is defined as the standard deviation of the 
results of all repetitions. 

For our analysis, we do not include the errors derived from the convariance matrix. The primary reason is that the flux errors of each 
wavelength element that are required for the precise calculation of the convariance matrix, unfortunately, are not provided 
for our spectra. Therefore, in such cases one has to be careful with the interpretation of the values of the covariance matrix. 

Table~\ref{table1} shows the errors derived from the two different sources described above. The errors in $T_{\mathrm{eff}}$ and 
$\log g$ due to the different initial parameters are slightly more significant whereas for $[Fe/H]$ and $\upsilon\sin i$ both 
type of errors are comparable. Finally, we add quadratically the 2 sources of errors that are described above (see Table~\ref{table2}). 
We notice that for higher $\upsilon\sin i$, the uncertainties in all parameters become higher as one would expect. K-type stars 
have also higher uncertainties compared to F- and G-types. In particular, the uncertainties in $\upsilon\sin i$, for K-type stars, 
are significantly high for $\upsilon\sin i>$\,45\,km/s. Fortunately, K-type stars are typically low rotators since rotational velocity 
decreases with the spectral type for FGK stars \citep[e.g.][]{gray1984, nielsen2013}.  

%These errors will give an estimation of how homogeneous can our sample be in case of using different wavelength intervals.
%SME uses a fixed error on the spectral flux for a fixed signal-to-noise (S/N) of 100. In order to re-scale the errors, we 
%assume that the residuals purely come from the noise, i.e. the fit is perfect. With this hypothesis, the reduced $\chi^{2}$ is by 
%definition equal to unity. To make this determination, we assumed that the noise was the same at all wavelength points, and performed 
%the fit with an arbitrary value of S/N and re-scaled the errors returned by multiplying them by square root of $\chi^{2}$. 
%-----------------------------------------------------------------------
\begin{table*}
\begin{center}
\caption{Internal error analysis for each spectral type and different rotational velocities. Two sources of errors were 
considered: Initial parameters and Wavelength choice (dependence on the choice of the 90\% of the total intervals). Rotational velocity 
0\,km/s refers to the intrinsic $\upsilon\sin i$ of the star.}
\label{table1}
%\scalebox{0.8}{
\begin{tabular}{lcccccc|cccccc}
\hline\hline
Parameters & \multicolumn{12}{c}{F-type (HD\,61421)} \\
\hline
%    \cline{2-6} \cline{7-10} \cline{11-13}
 & \multicolumn{6}{c}{Initial parameters} & \multicolumn{6}{c}{Wavelength choice} \\
$\upsilon\sin i$ & 0 km/s & 15 km/s & 25 km/s & 35 km/s & 45 km/s & 55 km/s & 0 km/s & 15 km/s & 25 km/s & 35 km/s & 45 km/s & 55 km/s \\
\hline
$T_{\mathrm{eff}}$ (K) & 27 & 43 & 48 & 54 & 97 & 108 & 13 & 9 & 17 & 48 & 16 & 85 \\
$\log g$ (dex) & 0.11 & 0.15 & 0.21 & 0.25 & 0.20 & 0.17 & 0.02 & 0.02 & 0.08 & 0.16 & 0.03 & 0.05  \\
$[Fe/H]$ (dex) & 0.03 & 0.04 & 0.05 & 0.07 & 0.06 & 0.07 & 0.01 & 0.01 & 0.02 & 0.04 & 0.01 & 0.01 \\
$\upsilon\sin i$ (km/s) & 0.20 & 0.25 & 0.98 & 1.00 & 1.57 & 2.60 & 0.08 & 0.14 & 0.91 & 0.89 & 0.93 & 2.13 \\
\hline
 & \multicolumn{12}{c}{G-type (Sun)} \\
    \cline{2-13} 
 & \multicolumn{6}{c}{Initial parameters} & \multicolumn{6}{c}{Wavelength choice} \\
$\upsilon\sin i$ & 0 km/s & 15 km/s & 25 km/s & 35 km/s & 45 km/s & 55 km/s & 0 km/s & 15 km/s & 25 km/s & 35 km/s & 45 km/s & 55 km/s \\
\hline
$T_{\mathrm{eff}}$ (K) & 12 & 5 & 18 & 86 & 86 & 147 & 13 & 9 & 21 & 36 & 43 & 116 \\
$\log g$ (dex) & 0.06 & 0.05 & 0.09 & 0.16 & 0.11 & 0.14 & 0.01 & 0.03 & 0.07 & 0.07 & 0.07 & 0.12 \\
$[Fe/H]$ (dex) & 0.02 & 0.01 & 0.04 & 0.07 & 0.07 & 0.07 & 0.02 & 0.01 & 0.03 & 0.02 & 0.02 & 0.04 \\
$\upsilon\sin i$ (km/s) & 0.25 & 0.09 & 0.32 & 0.36 & 2.47 & 3.67 & 0.11 & 0.99 & 0.26 & 0.17 & 2.29 & 3.49 \\ 
\hline
 & \multicolumn{12}{c}{K-type (HD\,20868)} \\
\cline{2-13} 
 & \multicolumn{6}{c}{Initial parameters} & \multicolumn{6}{c}{Wavelength choice} \\
$\upsilon\sin i$ & 0 km/s & 15 km/s & 25 km/s & 35 km/s & 45 km/s & 55 km/s & 0 km/s & 15 km/s & 25 km/s & 35 km/s & 45 km/s & 55 km/s \\
\hline
$T_{\mathrm{eff}}$ (K) & 20 & 52 & 54 & 77 & 130 & 127 & 15 & 47 & 45 & 68 & 110 & 110 \\
$\log g$ (dex) & 0.09 & 0.16 & 0.17 & 0.20 & 0.14 & 0.18 & 0.03 & 0.11 & 0.12 & 0.16 & 0.10 & 0.09 \\
$[Fe/H]$ (dex) & 0.02 & 0.06 & 0.09 & 0.09 & 0.09 & 0.09 & 0.02 & 0.05 & 0.07 & 0.07 & 0.02 & 0.01 \\
$\upsilon\sin i$ (km/s) & 0.46 & 0.72 & 0.86 & 2.38 & 6.92 & 7.54 & 0.45 & 0.69 & 0.77 & 2.12 & 6.98 & 7.36 \\ 
\hline
\end{tabular}
%}
\end{center}
\end{table*}
%-----------------------------------------------------------------

%-----------------------------------------------------------------------
  \begin{table*}
     \centering
     \caption[]{Errors summed quadratically for each spectral type and for the different rotational velocities.}
     \label{table2}
        \begin{tabular}{lcccccc}
           \hline\hline
Parameters & \multicolumn{6}{c}{F-type (HD\,61421)} \\
\hline
%    \cline{2-6} \cline{7-10} \cline{11-13}
$\upsilon\sin i$ & 0 km/s & 15 km/s & 25 km/s & 35 km/s & 45 km/s & 55 km/s \\
\hline
$T_{\mathrm{eff}}$ (K) & 30 & 44 & 51 & 72 & 98 & 137 \\
$\log g$ (dex) & 0.11 & 0.15 & 0.22 & 0.30 & 0.20 & 0.18 \\
$[Fe/H]$ (dex) & 0.03 & 0.04 & 0.05 & 0.03 & 0.08 & 0.06 \\
$\upsilon\sin i$ (km/s) & 0.22 & 0.29 & 1.34 & 1.34 & 1.82 & 3.36 \\
\hline
& \multicolumn{6}{c}{G-type (Sun)} \\
 \cline{2-7}
$\upsilon\sin i$ & 0 km/s & 15 km/s & 25 km/s & 35 km/s & 45 km/s & 55 km/s \\
\hline
$T_{\mathrm{eff}}$ (K) & 18 & 10 & 28 & 93 & 96 & 187 \\
$\log g$ (dex) & 0.06 & 0.06 & 0.11 & 0.17 & 0.13 & 0.18 \\
$[Fe/H]$ (dex) & 0.03 & 0.01 & 0.05 & 0.07 & 0.07 & 0.08 \\
$\upsilon\sin i$ (km/s) & 0.27 & 0.99 & 0.41 & 0.40 & 3.37 & 5.06 \\
\hline
& \multicolumn{6}{c}{K-type (HD\,20868)} \\
 \cline{2-7}
$\upsilon\sin i$ & 0 km/s & 15 km/s & 25 km/s & 35 km/s & 45 km/s & 55 km/s \\
\hline
$T_{\mathrm{eff}}$ (K) & 25 & 70 & 70 & 103 & 170 & 168 \\
$\log g$ (dex) & 0.09 & 0.19 & 0.21 & 0.26 & 0.17 & 0.20 \\
$[Fe/H]$ (dex) & 0.03 & 0.08 & 0.11 & 0.11 & 0.09 & 0.09 \\
$\upsilon\sin i$ (km/s) & 0.64 & 0.99 & 1.15 & 3.19 & 9.83 & 10.54 \\
\hline
\end{tabular}
  \end{table*}

\section{Spectroscopic parameters for low rotating FGK stars}\label{results1}

\begin{figure}[t!]
  \centering
   \includegraphics[width=1.0\linewidth, height=7.0cm]{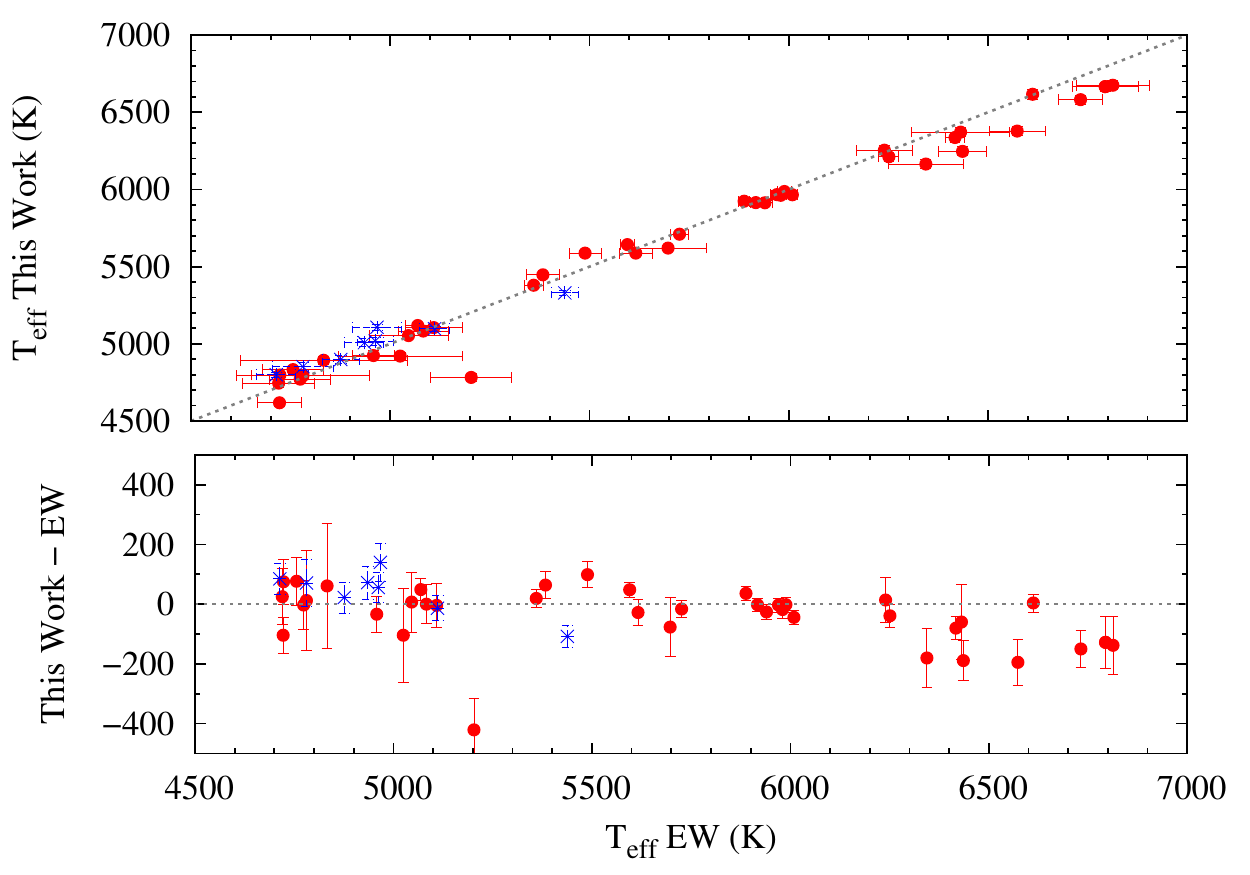}
   \includegraphics[width=1.0\linewidth, height=7.0cm]{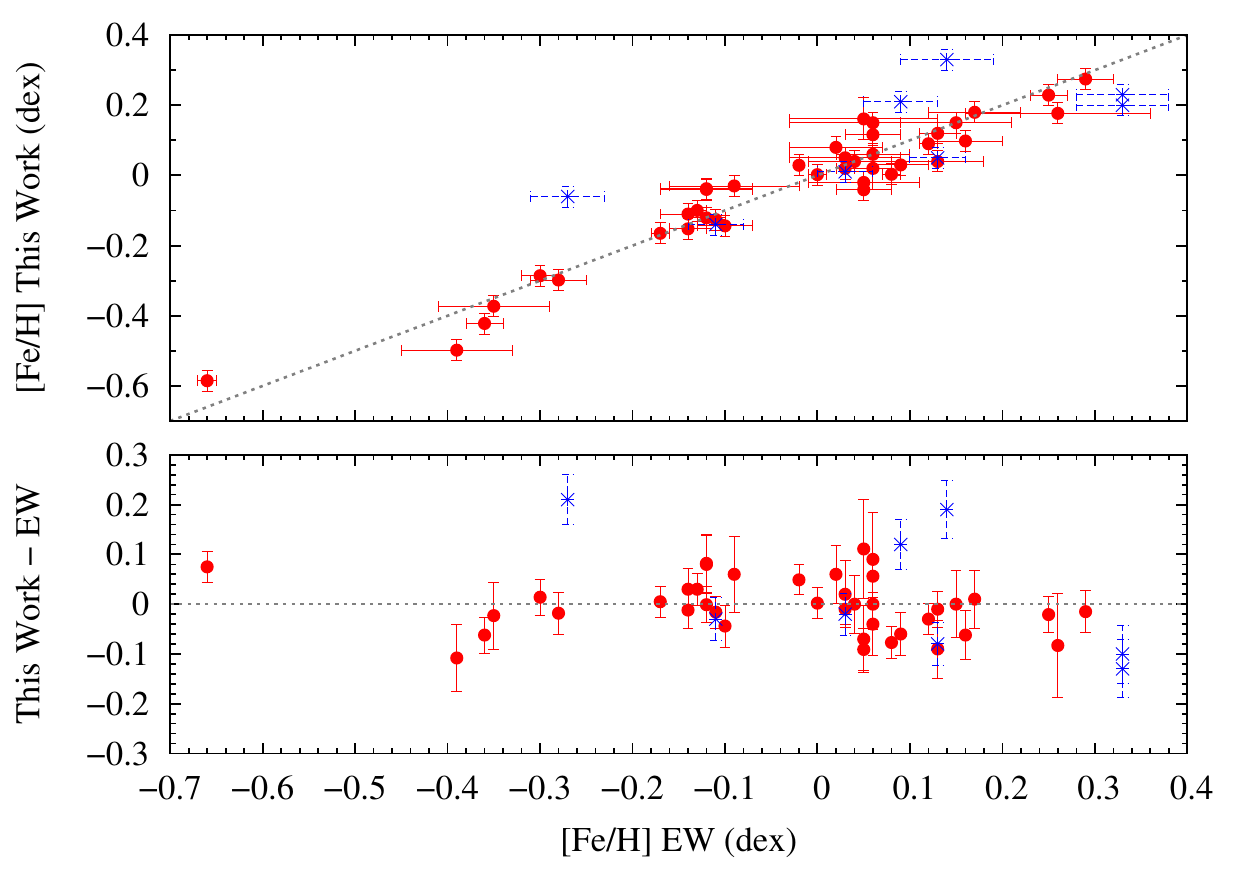}
   \includegraphics[width=1.0\linewidth, height=7.0cm]{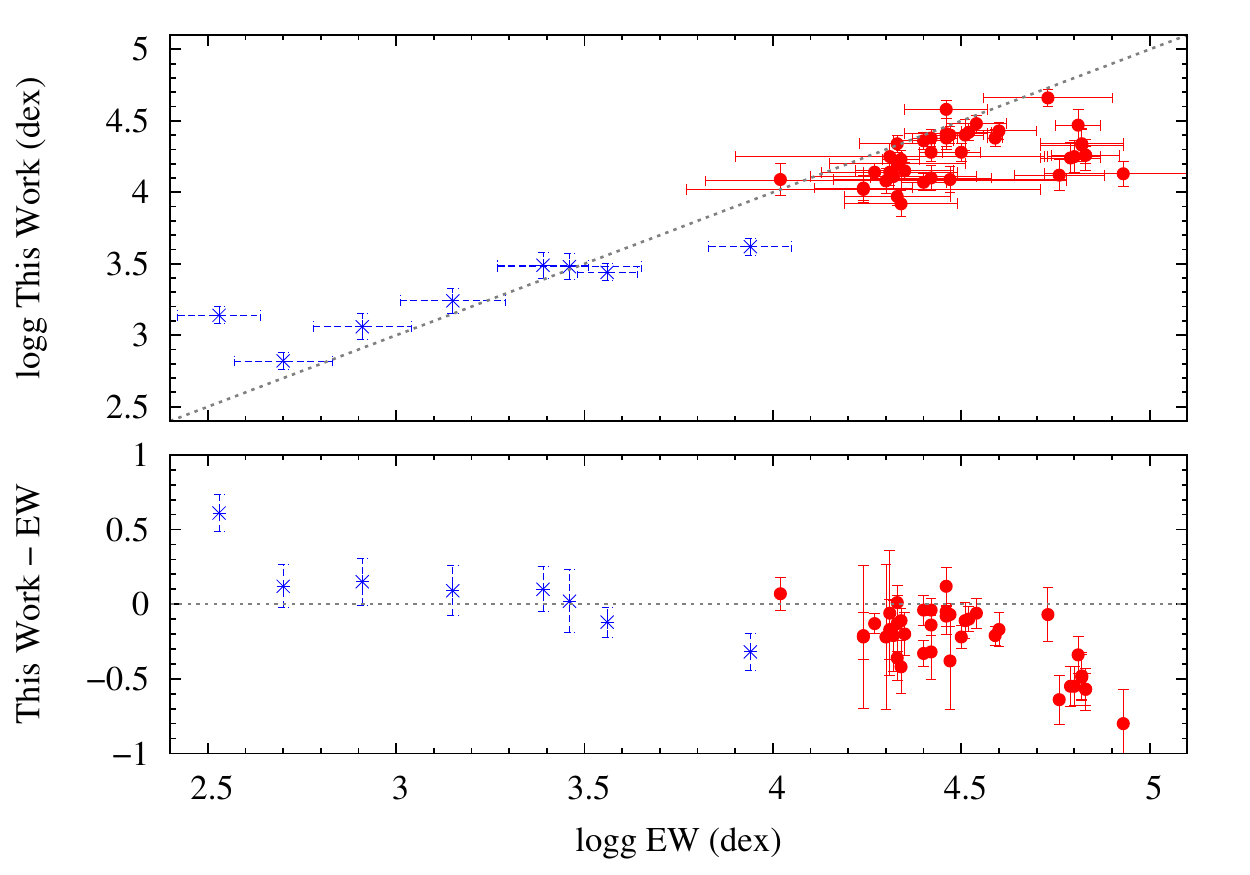}
  \caption{\footnotesize{Comparison between the parameters derived using the spectral synthesis method (This Work) and the results 
of our EW method: temperature (top panel), metallicitity (middle panel) and surface gravity (bottom panel). Circles represent dwarf 
stars and asterisks giants.}}
  \label{fig:2}
  \end{figure}

To test the effectiveness of the line list, we use a sample of 48 FGK stars (40 dwarfs and 8 giants) with slow rotation, 
high S/N and high resolution spectra, most of them taken from the archival data of HARPS (R\,$\sim$\,110000) and the 
rest with UVES (R\,$\sim$\,110000) and FEROS (R\,$\sim$\,48000) spectrographs. 
Their stellar parameters range from: 4758\,$\leq T_{\mathrm{eff}} \leq$\,6666\,K, 2.82\,$\leq \log g \leq$\,4.58\,dex, 
and -0.58\,$\leq$\,[Fe/H]\,$\leq$\,0.33\,dex and are derived following the method described in Sect.~\ref{method}. 
Figure~\ref{fig:2} depicts the comparison between the parameters derived in this work and the EW method. All parameters from 
the EW method were taken from \cite{sousa3,mortier13,mortier13_gravity,tsantaki13,santos13} using the same methodology that 
provides best possible homogeneity for the comparison. The differences between these methods are presented in Table~\ref{table3} 
and the stellar parameters in Table~\ref{table8}.      

%-----------------------------------------------------------------------
  \begin{table*}
     \centering
     \caption[]{Differences in stellar parameters between this work and the EW method for the 48 sample stars. MAD 
values correspond to the median average deviation and are indicated in parenthesis. The differences are also listed per 
spectral type. N is the number of stars used for the comparison.}
     \label{table3}
        \begin{tabular}{ccccc}
           \hline\hline
This Work -- EW  & $\Delta T_{\mathrm{eff}}$\tablefootmark{a} ($\pm$\,MAD)\,K & $\Delta \log g$\,($\pm$\,MAD)\,dex & $\Delta$ [Fe/H]\,($\pm$\,MAD)\,dex & N \\
\hline
Whole sample     & -26 $\pm$ 14 ($\pm$ 55) & -0.19 $\pm$ 0.04 ($\pm$ 0.14) & 0.000 $\pm$ 0.010 ($\pm$ 0.041) & 48 \\
%                       &  ($\sigma$ = 98) & ($\sigma$ = 0.25) & ($\sigma$ = 0.07) \\
F-type      & -97 $\pm$ 22 ($\pm$ 68) & -0.34 $\pm$ 0.07 ($\pm$ 0.18) & 0.006 $\pm$ 0.014 ($\pm$ 0.014) & 12 \\
G-type      & 7 $\pm$ 16 ($\pm$ 36)   & -0.07 $\pm$ 0.05 ($\pm$ 0.04) & 0.019 $\pm$ 0.021 ($\pm$ 0.032) & 18 \\
K-type      & -5 $\pm$ 27 ($\pm$ 32)  & -0.18 $\pm$ 0.05 ($\pm$ 0.16) & -0.027 $\pm$ 0.016 ($\pm$ 0.042) & 18\\
         \hline
        \end{tabular}
\tablefoot{\tablefoottext{a}{The standard errors of the mean ($\sigma_{M}$) are calculated with the 
following formula: $\sigma_{M}$=$\frac{\sigma}{\sqrt{N}}$, $\sigma$ being the standard deviation.}}
  \end{table*}
%-------------------------------------------------------------------------

%----------------------------------------- FIGURE ---------------------------------------------------
%--------------------------------------------------------------------------------------------------

The effective temperatures derived with the spectral synthesis technique and the EW method are in good agreement. The greatest 
discrepancies appear for $T_{\mathrm{eff}}>$\,6000\,K, where the effective temperature derived from this work is systematically cooler. 
The same systematics are also presented in \cite{molenda13}, where the authors compare the EW method with other spectral synthesis 
techniques but the explanation for these discrepancies is not yet clear.

The values of metallicitity are in perfect agreement between the two methods. 

Surface gravity is a parameter that is the most difficult to constrain with spectroscopy. The comparison of the two methods shows 
a considerable offset of 0.19\,dex, where $\log g$ is underestimated compared to the EW method. Interestingly, this offset is smaller for 
giant stars ($\Delta\log g$\,=\,0.07\,dex) than for dwarfs ($\Delta\log g$\,=\,-0.24\,dex).   

%----------------------------------------- FIGURE ---------------------------------------------------
\begin{figure}
  \centering
   \includegraphics[width=1.0\linewidth]{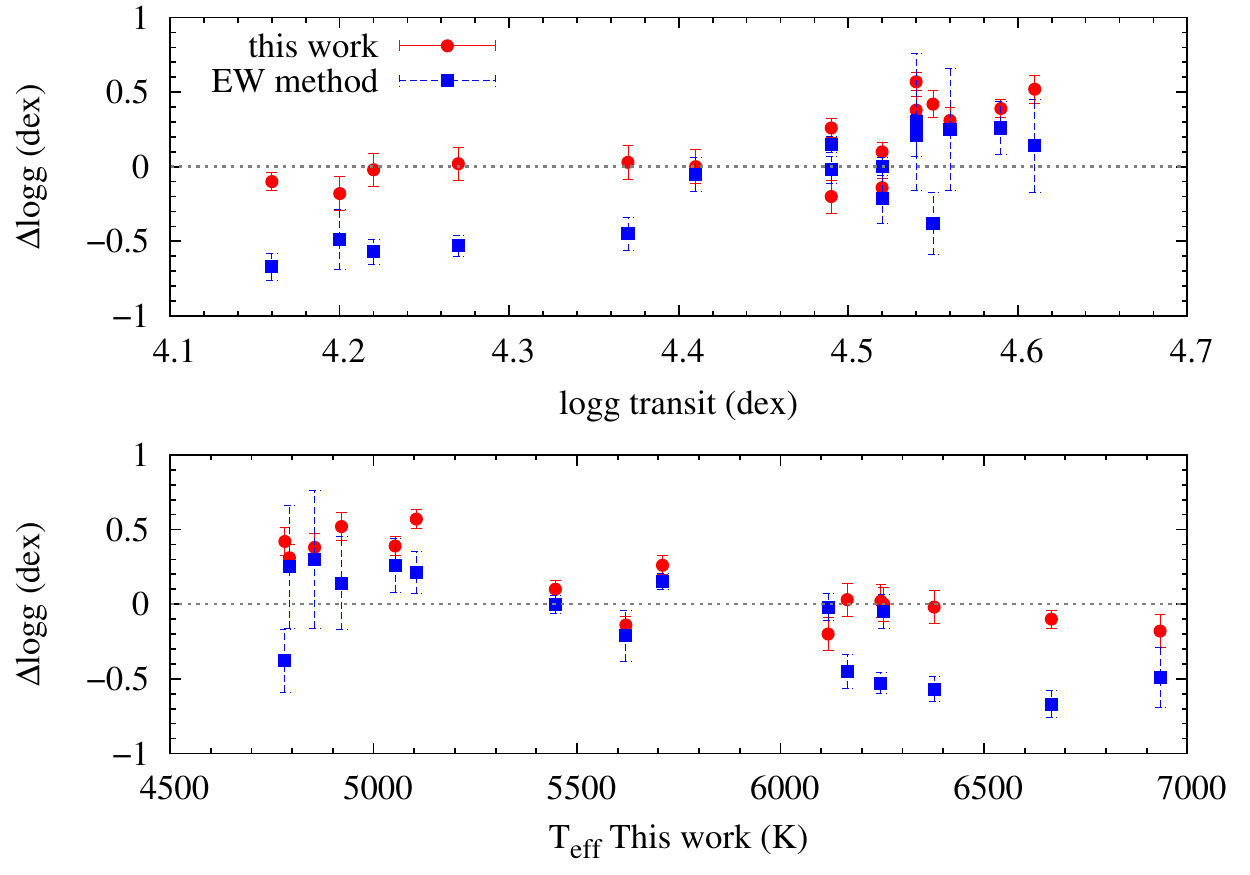}
  \caption{\footnotesize{Comparison of surface gravity derived from the transit fit with this work and the EW method. $\Delta \log g$ 
represents 'transit minus this work' (red circles) and 'transit minus EW method' (blue squares).}}
  \label{low_transit_logg}
  \end{figure}
%--------------------------------------------------------------------------------------------------

To further investigate these differences, we compare the spectroscopic $\log g$ with surface gravity derived with another method that is 
less model dependent. For 16 dwarf stars in our sample that have a transiting planet, surface gravity can be derived from the analysis of 
the transit light curve (see also Sect.\,5). We compare $\log g$ derived from the transit light curve with the spectroscopic $\log g$ 
from the EW method (both values are taken from \cite{mortier13_gravity}) and this work (see Fig.~\ref{low_transit_logg}). 

We show that $\log g$ from the EW analysis is overestimated for low $\log g$ values and underestimated for high $\log g$ values. 
Fortunately, this trend does not affect $T_{\mathrm{eff}}$ and $[Fe/H]$ as shown in the recent work of \cite{torres12}. 
%However, attention should be paid when using this determination of $\log g$ to distinguish stars as dwarfs or (sub-)giants as the 
%differences are $\sim$0.40\,dex for the more evolved stars. 
Same systematics were also found between the $\log g$ from the EW method and the $\log g$ derived with the \textit{Hipparcos} parallaxes 
for solar-type stars in \cite{tsantaki13} and \cite{bensby2014}.
%                       &  ($\sigma$ = 98) & ($\sigma$ = 0.25) & ($\sigma$ = 0.07) \\
These results imply that $\log g$ from the EW method using iron lines suffers from biases, but there is no clear explanation for the 
reasons. 

On the other hand, $\log g$ derived from this work is in very good agreement with the transit $\log g$, for values lower than 4.5\,dex. 
Stars with $\log g>$\,4.5\,dex correspond to the cooler stars and are also underestimated. 
%Since the calculation of $\log g$ with the spectral synthesis technique does not involve ionization equilibrium assumptions for iron 
%lines as for the EW method, this underestimation does not imply deviations from the ionization equilibrium for cool stars. 
The reason for this underestimation is not yet known and further investigation is required to understand this behaviour.

Despite the differences for the $\log g$ values of mainly the F-type stars, the results listed in Table~\ref{table3} 
show that for low rotating FGK stars, stellar parameters derived from both methods are on the same scale. This means that 
for the whole sample the residuals between both methods are small and of the same order of magnitude as are the errors of 
the parameters.

\section{Spectroscopic parameters for high rotating FGK dwarfs}\label{rotation}

Testing our method for low FGK rotators does not necessary imply that it will work for higher $\upsilon\sin i$ where spectral lines 
are much broadened and shallower. Our goal is to examine how efficient our method is for moderate/high rotating stars. For this 
purpose, we derive stellar parameters for reference stars of different spectral type and with low $\upsilon\sin i$. Secondly, these 
stars are convolved with a set of rotational profiles using the \texttt{rotin3} routine as part of the SYNSPEC synthesis 
code\footnote{http://nova.astro.umd.edu/Synspec43/} \citep{rotin3}. As a result, each star has 8 different rotational velocities 
(initial, 5, 10, 15, 20, 25, 30, 40, 50\,km/s). Stellar parameters of all rotational profiles are calculated to investigate how 
they differ from the non-broadened (unconvolved) star. This test is an indication of how the accuracy of our method is affected by adding 
a rotational profile. 

\begin{figure}
  \centering
   \includegraphics[width=1.0\linewidth]{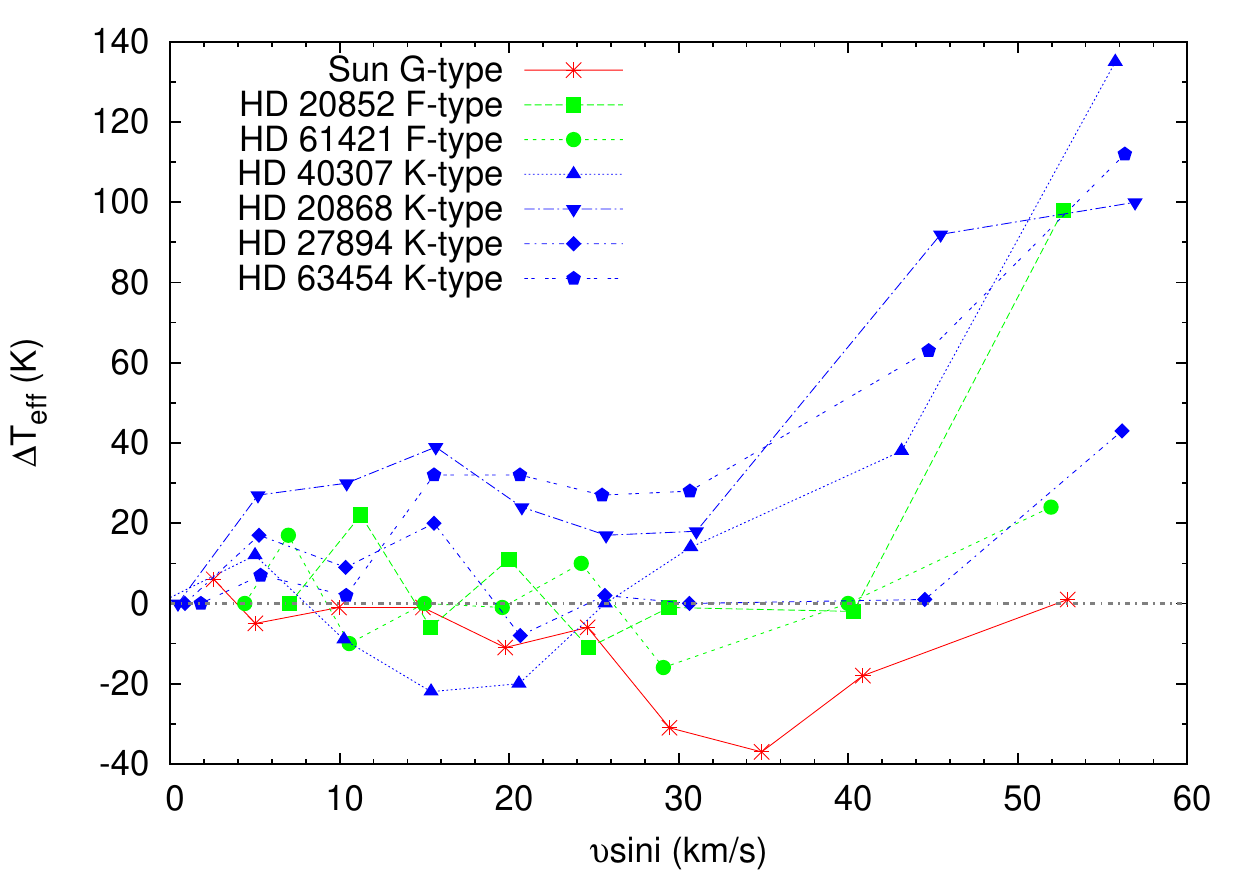}
   \includegraphics[width=1.0\linewidth]{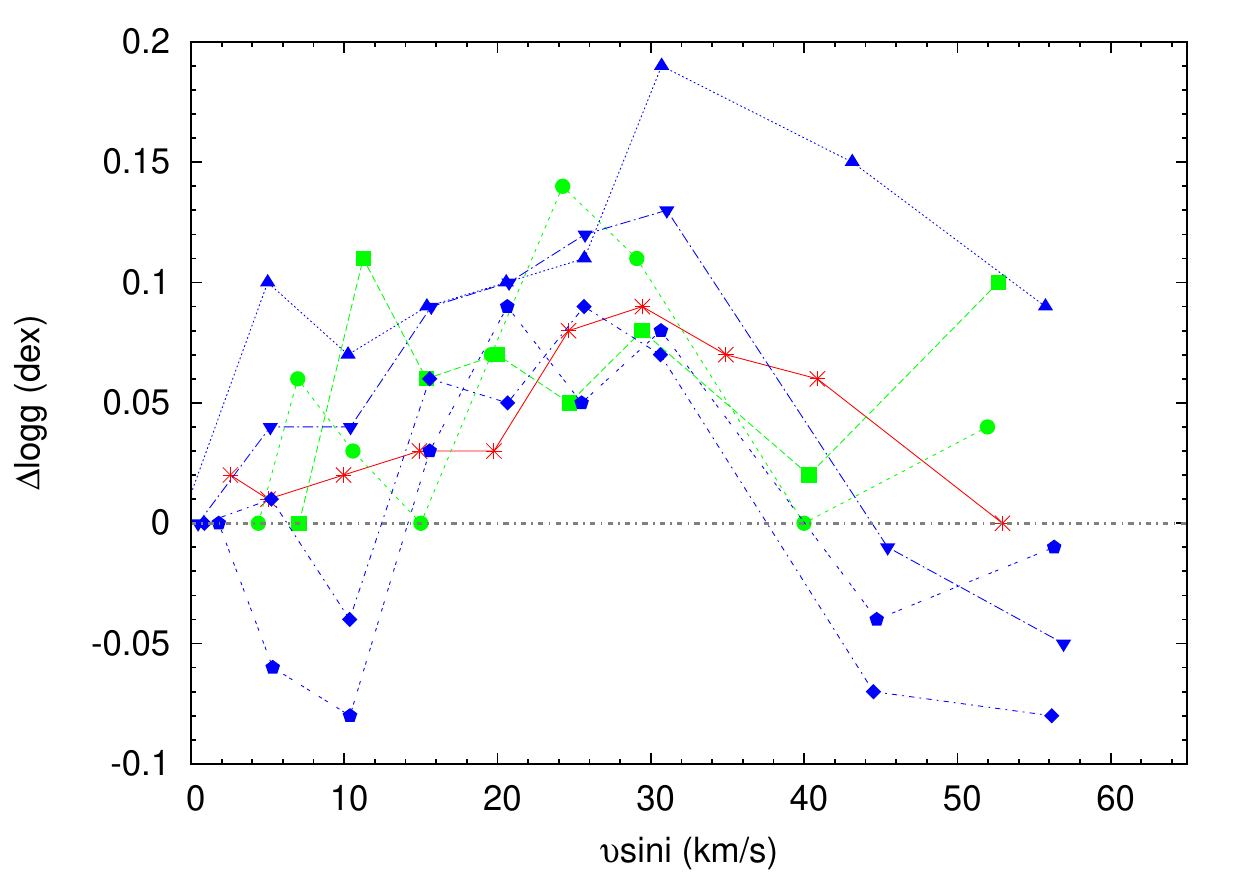}
   \includegraphics[width=1.0\linewidth]{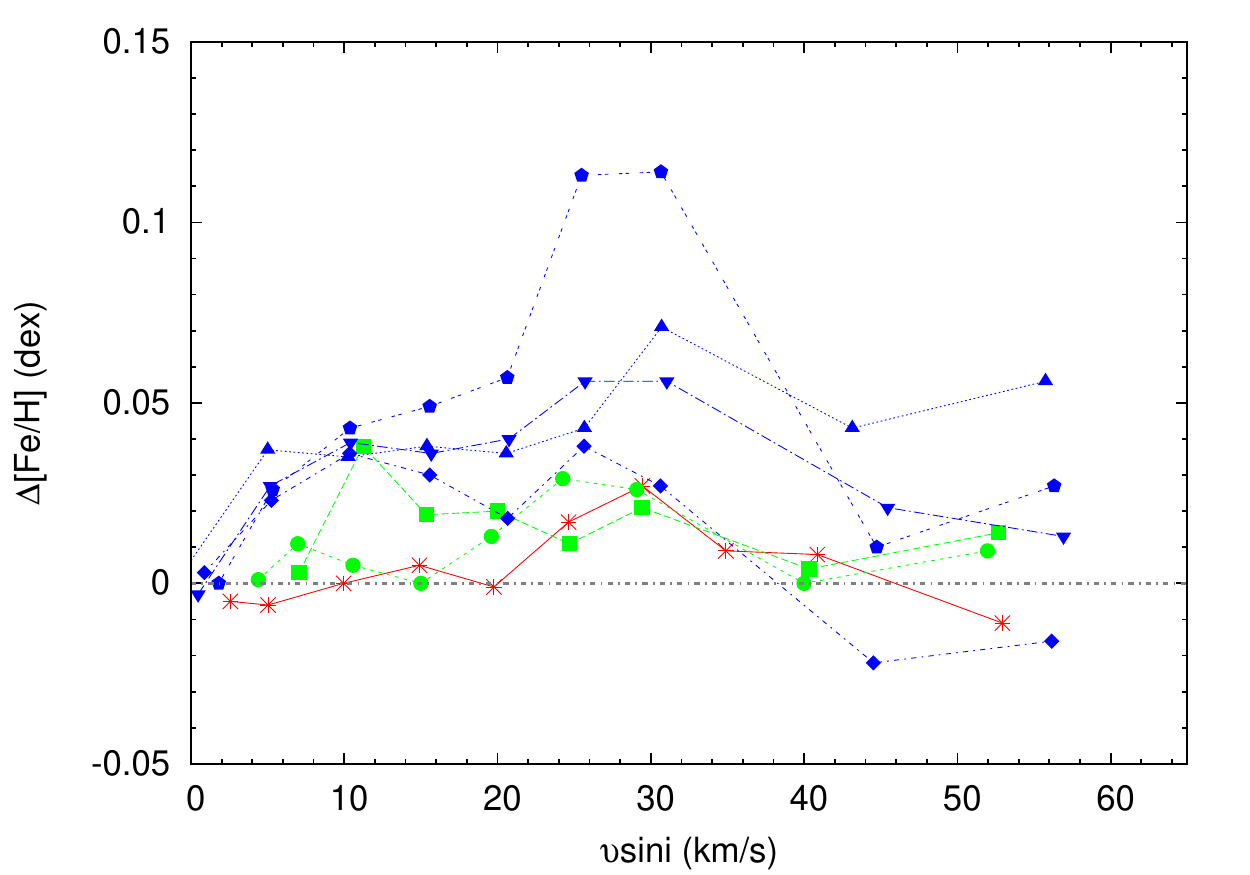}
  \caption{\footnotesize{Differences in temperature (top panel), surface gravity (middle panel) and metallicitity (bottom panel) 
(initial $\upsilon\sin i$ minus the different rotational profiles) vs. $\upsilon\sin i$. Each star is represented with different 
symbol and each spectral type is represented with different colour. Blue for K-type, red for G-type and green F-type. The initial 
$\upsilon\sin i$ is different for each star.}}
  \label{fig:3}
  \end{figure}

The selected reference stars are: two F-type, one G-type, and four K-type stars and are presented in boldface in Table~\ref{table8}. 
Probably one star per spectral type would be enough but we included more F- and K-type stars because they showed higher uncertainties 
(especially the K-type stars). In Fig.~\ref{fig:3} the differences of stellar parameters between the stars with the 
unconvolved values (original $\upsilon\sin i$) and the convolved ones are plotted for the 8 different rotational velocities.  

%----------------------------------------- FIGURE ---------------------------------------------------
%--------------------------------------------------------------------------------------------------

%----------------------------------------- FIGURE ---------------------------------------------------
\begin{figure}
  \centering
   \includegraphics[width=1.0\linewidth]{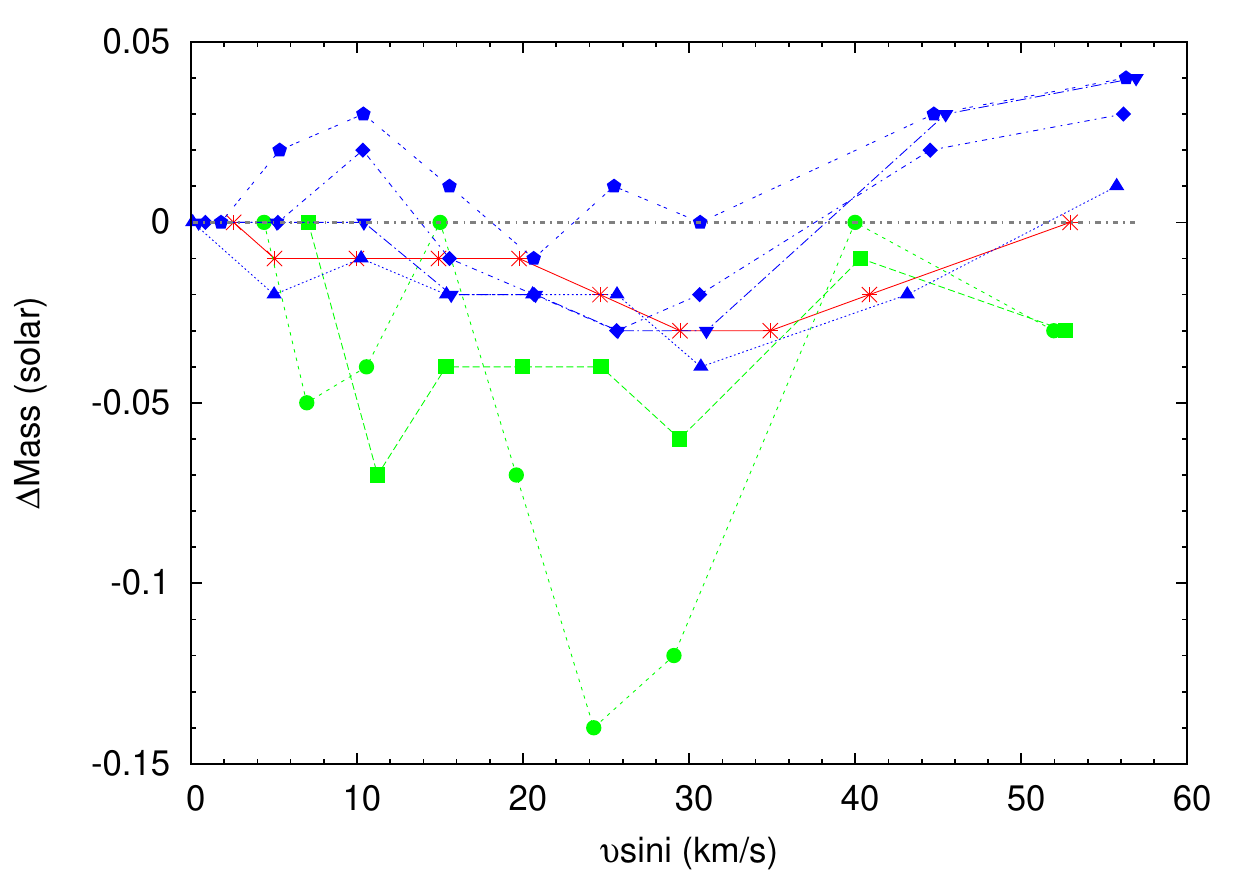}
   \includegraphics[width=1.0\linewidth]{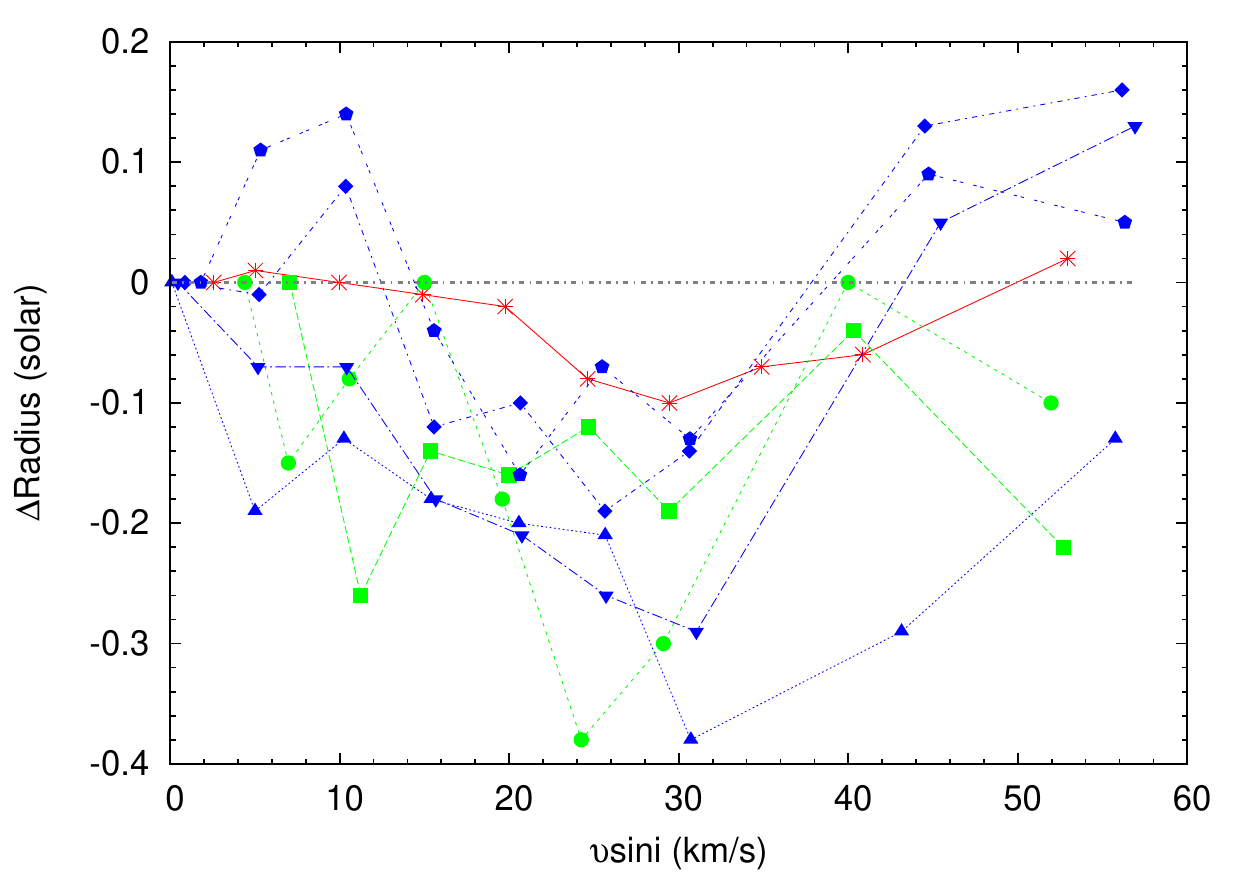}
  \caption{\footnotesize{Differences in stellar mass (top panel) and radius (bottom panel) vs. $\upsilon\sin i$. The 
symbols are the same as in Fig.~\ref{fig:3}.} }
  \label{mass_radius_rot}
  \end{figure}
%--------------------------------------------------------------------------------------------------

As $\upsilon\sin i$ increases, K-type stars show the highest differences in the stellar parameters compared to the non-broadened 
profile. These deviations for high $\upsilon\sin i$ are also shown in the error analysis of Sect.~\ref{error_analysis}. The temperatures 
of these stars are systematically underestimated with increasing $\upsilon\sin i$. On the other hand, the 
parameters of F- and G-type stars are very close to the ones with slow rotation and no distinct trends are observed with rotation. Even for 
very high $\upsilon\sin i$, temperature and metallicitity can be derived with differences in values of less than 100\,K and 0.05\,dex 
respectively. Surface gravity, however, shows high differences that reach up to $\sim$0.20\,dex.

The above discrepancies in the parameters affect in turn the stellar mass and radius. To investigate these offsets, we 
calculate the mass and radius for all the rotational velocities using the calibration of \cite{torres10_mass} but corrected 
for small offsets to match masses derived from isochrone fits by \cite{santos13}. The results in Fig.~\ref{mass_radius_rot} 
show that the mass hardly changes as $\upsilon\sin i$ increases. Stellar radius however, is affected in the same manner as 
surface gravity with higher radius differences. For example, the maximum difference in $\log g$ ($\sim$0.20\,dex) causes a 
deviation in radius of 0.39\,R$_{\odot}$.

\subsection{Application to FGK high rotators}\label{high}

We select a sample of FGK dwarfs with moderate/high $\upsilon\sin i$, that have available several estimates of their parameters 
from different techniques in the literature (see references in Table\,A.2). From these references, we included 17 stars with 
$\upsilon\sin i$ up to 54\,km/s that have spectra available in the public archives of different high resolution instruments (HARPS, 
FEROS, ELODIE, and CORALIE). The spectra were already processed with their standard pipeline procedures. We corrected for the radial 
velocity shifts and in cases of multiple observations, the spectra are summed using the \texttt{IRAF} tools, \texttt{dopcor} and 
\texttt{scombine} respectively. 

The stellar parameters are derived with the method of this work and the results and literature properties of the sample are presented in 
Table\,A.2. Table~\ref{table5} shows the differences between the stellar parameters of this work and the different methods 
used: other spectral synthesis techniques, the EW method (until $\upsilon\sin i$ $\sim$ 10\,km/s) and the photometric technique, namely 
IRFM. The differences between this work and other methods are very small for all parameters.

\begin{figure}
  \centering
   \includegraphics[width=1.0\linewidth]{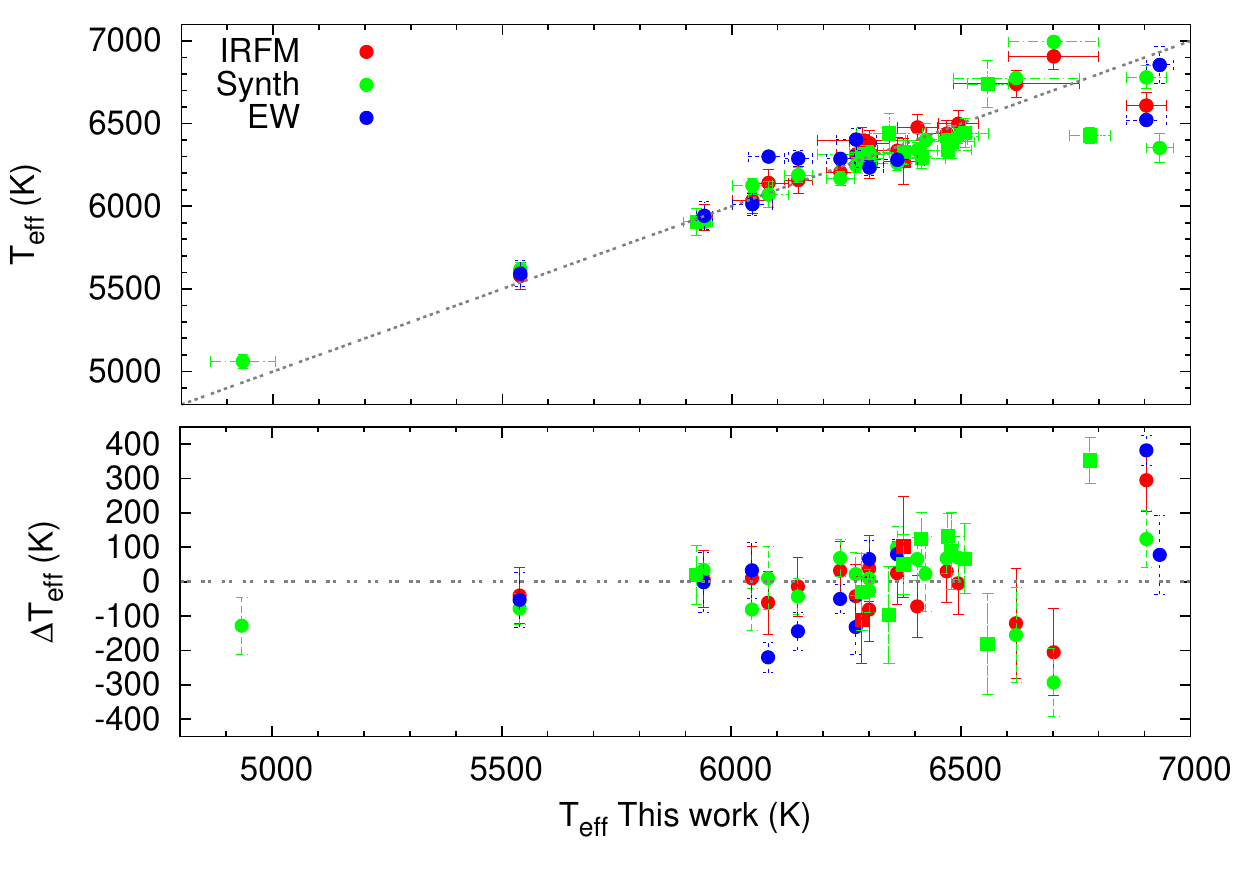}
   \includegraphics[width=1.0\linewidth]{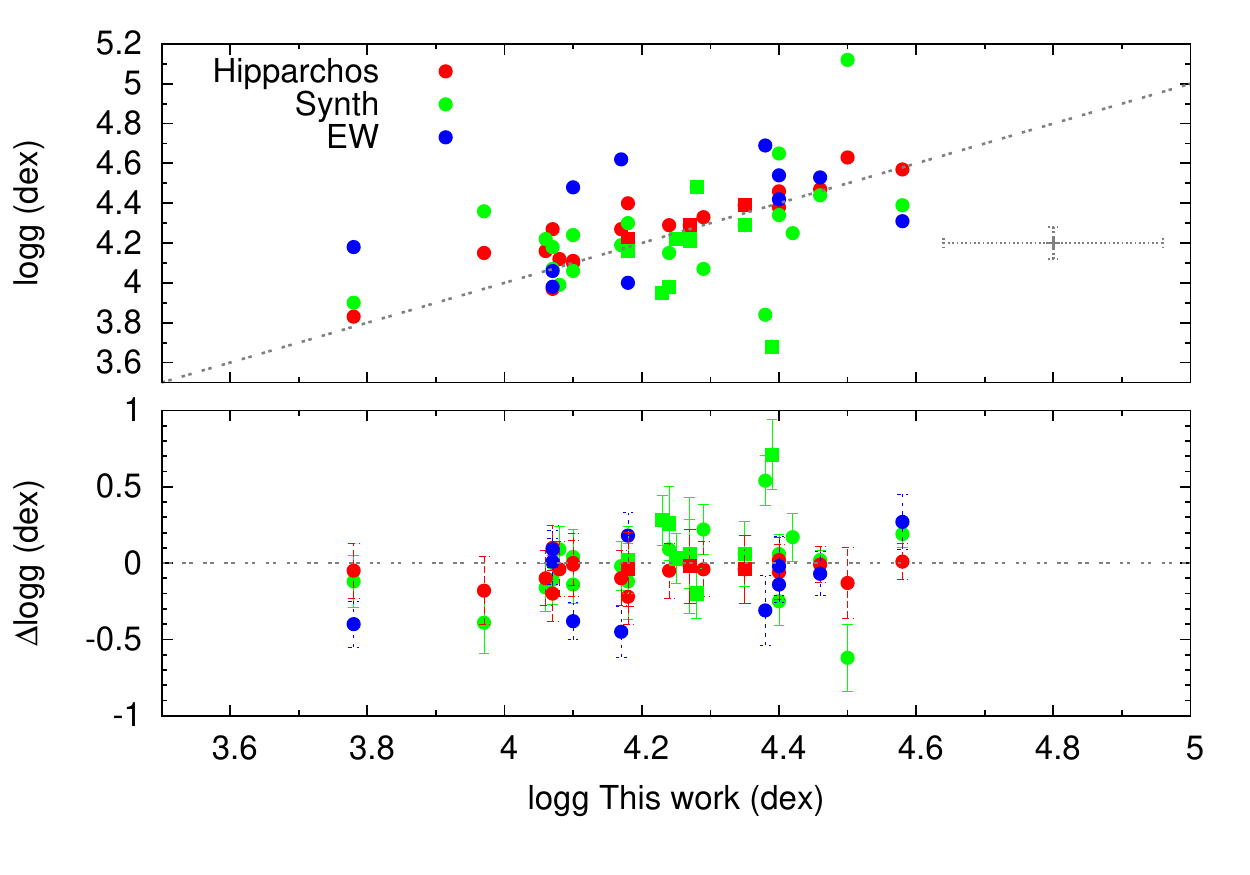}
   \includegraphics[width=1.0\linewidth]{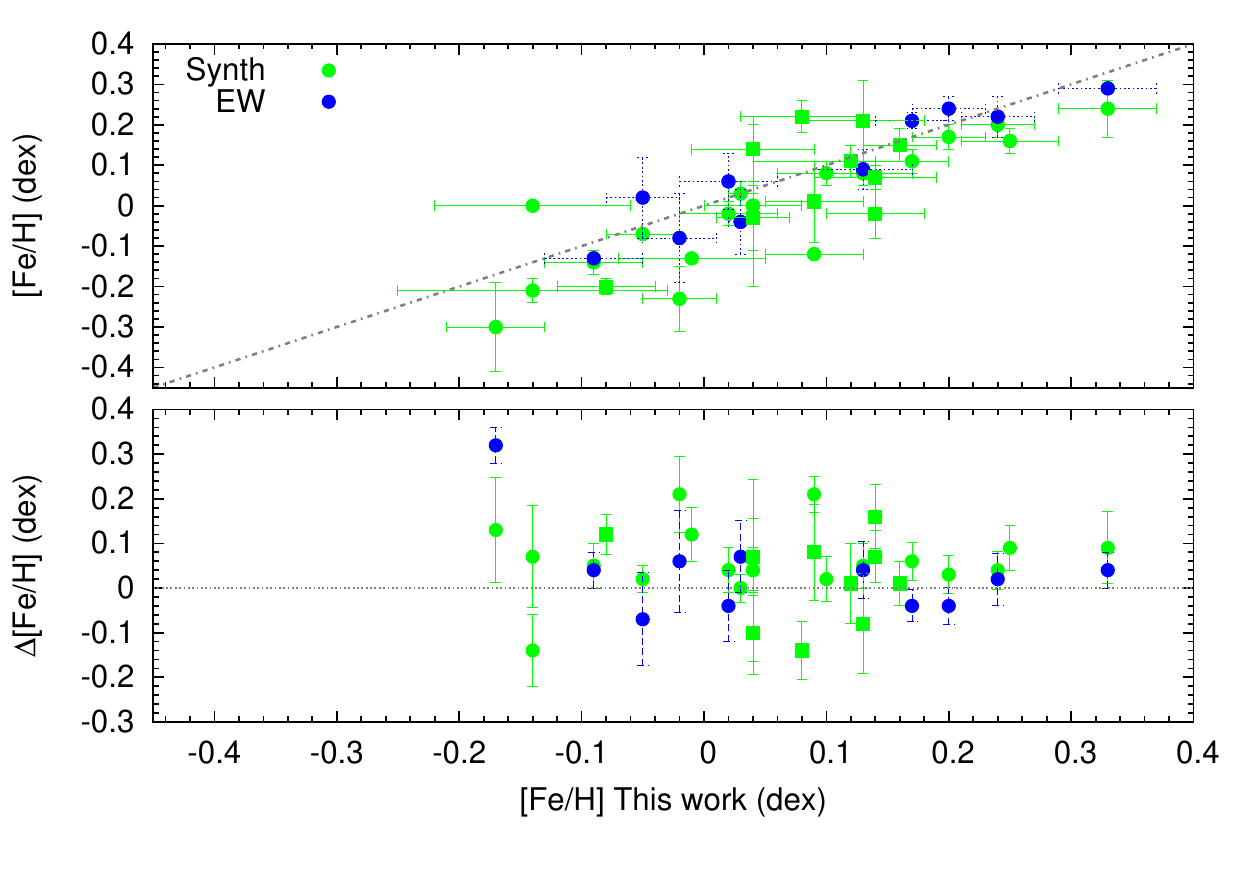}
  \caption{\footnotesize{Temperature (top panel), surface gravity (middle panel), and metallicitity (bottom panel). Different 
colors represent different techniques. Square symbols represent planet-hosts analyzed in this work. In the middle panel, the 
average error is plotted. In each panel, the upper plot compares the data and the lower plot compares the residual 
differences from perfect agreement.} }
  \label{fig:4}
  \end{figure}

\begin{figure}
  \centering
   \includegraphics[width=0.99\linewidth]{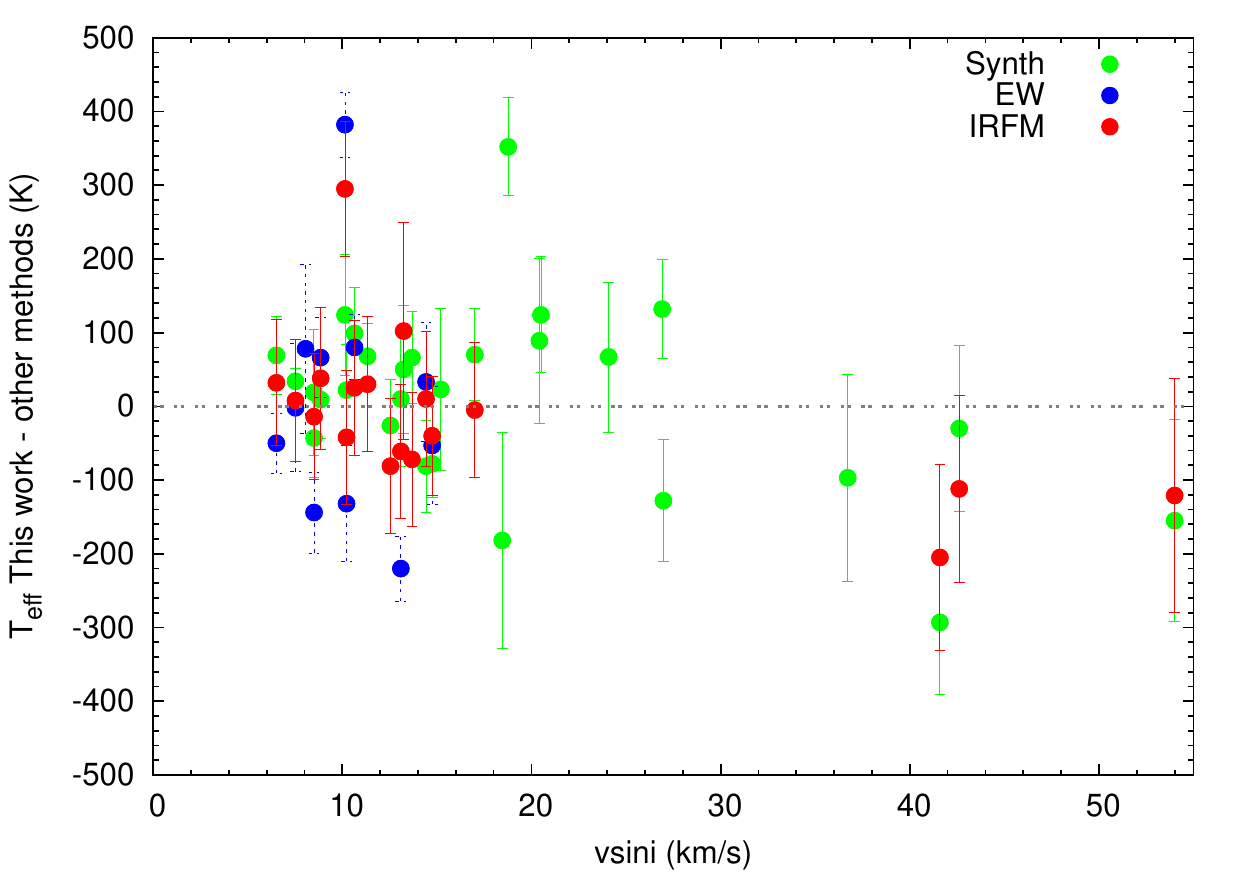}
   \includegraphics[width=0.99\linewidth]{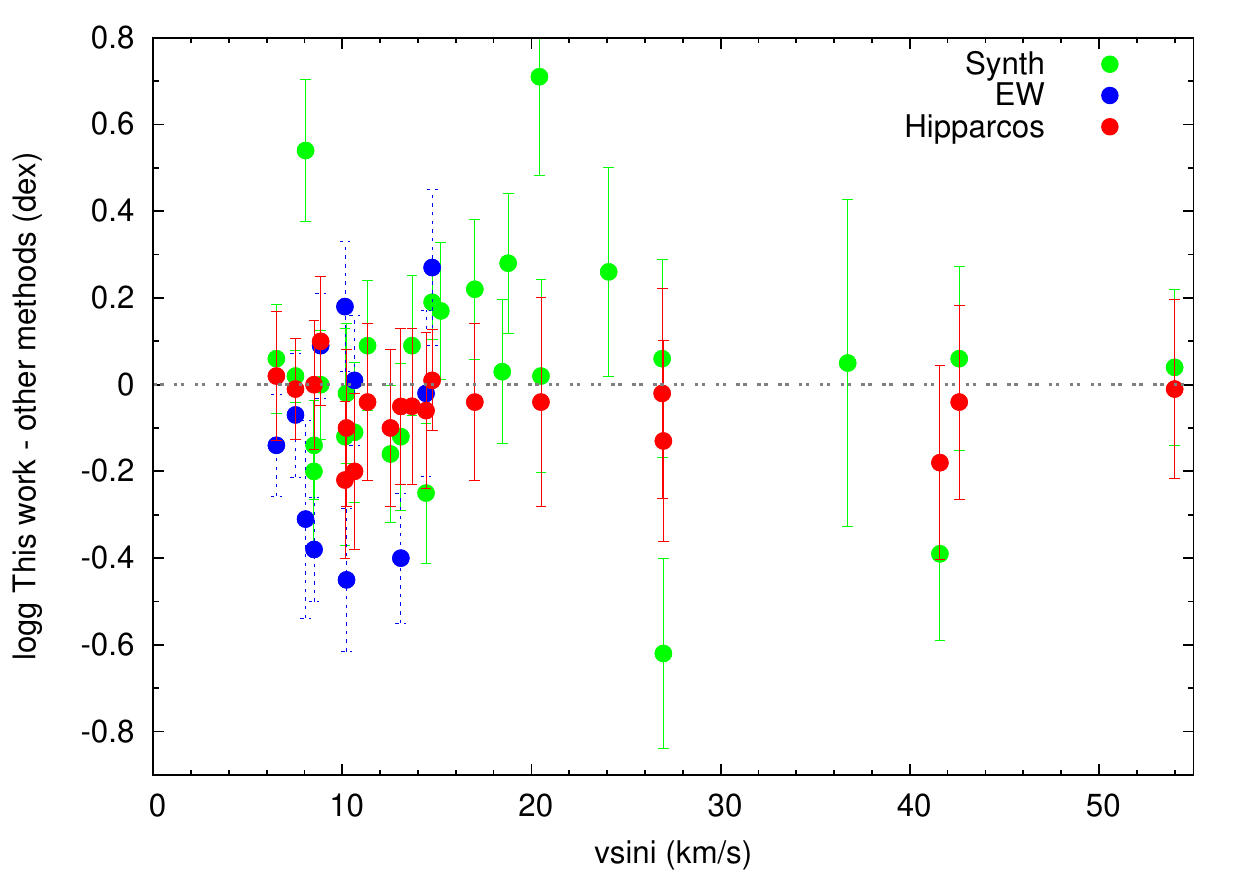}
   \includegraphics[width=0.99\linewidth]{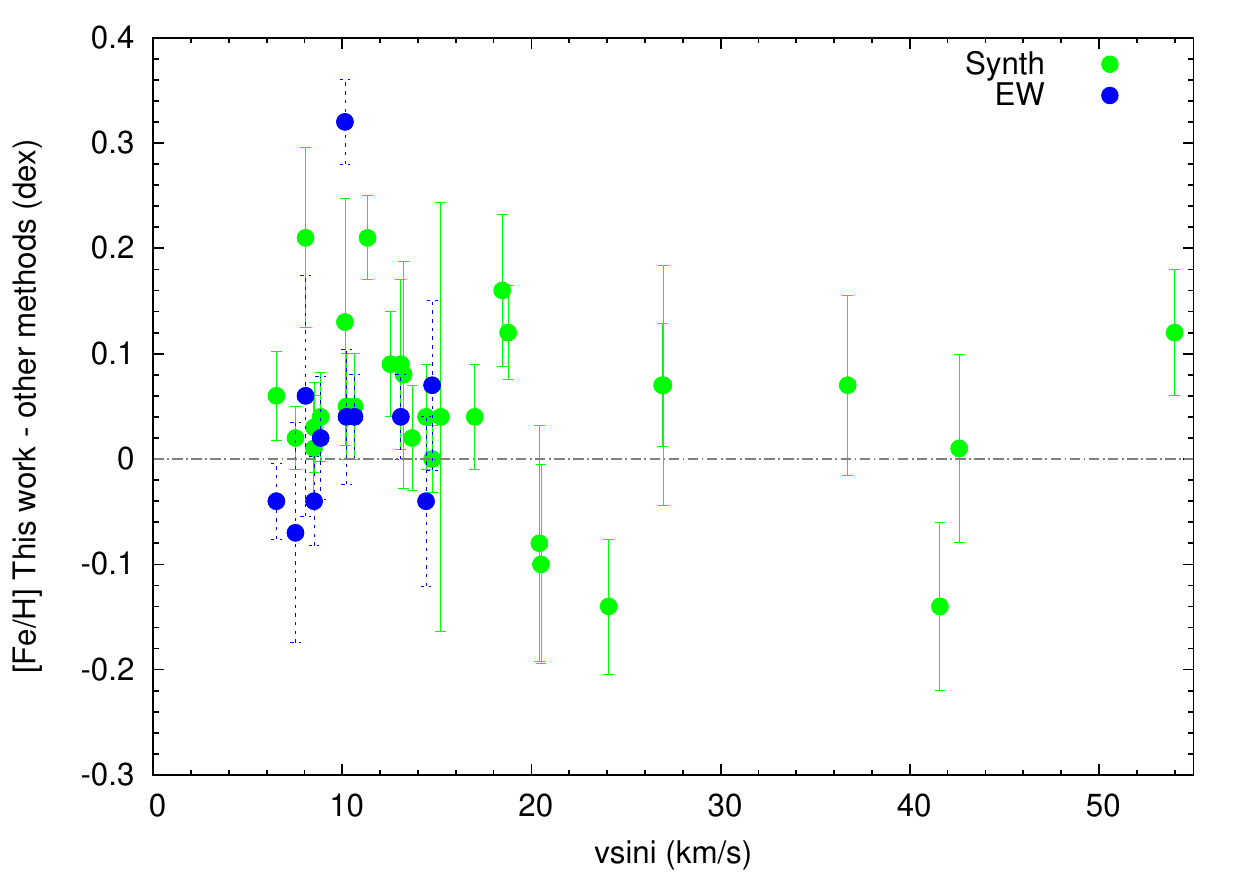}
  \caption{\footnotesize{Differences in temperature (top panel), surface gravity (middle panel) and metallicitity (bottom panel) 
versus rotational velocity for moderate/high rotators.}}
  \label{fig:5}
  \end{figure}

In Fig.~\ref{fig:4} we plot the comparison between literature values and our results. Figure~\ref{fig:5} shows the stellar parameters 
derived with different methods in dependence of rotational velocity from this work. Even though the mean differences in temperature are 
close to zero, there appears a slight overestimation of our method for high $\upsilon\sin i$. Surface gravity shows the lowest dispersion 
when compared to trigonometric $\log g$ from all methods. Metallicity is also in agreement, excluding perhaps an outlier (HD\,49933).

%-----------------------------------------------------------------------
\begin{table}
\caption{Differences in parameters derived with different methods. N indicates the number of stars used for the comparison.}
\label{table5}
\scalebox{0.87}{
\begin{tabular}{p{0.30\linewidth}ccccc}
           \hline\hline
           & $\Delta T_{\mathrm{eff}}$\,(K) & $\Delta \log g$\,(dex) & $\Delta$ [Fe/H]\,(dex) & N \\
\hline
\multirow{2}{*}{This Work -- EW}  & 3 $\pm$ 48 & -0.11 $\pm$ 0.07 & 0.04 $\pm$ 0.03 & 11 \\
%                       &  ($\sigma$ = 159) & ($\sigma$ = 0.25) & ($\sigma$ = 0.11) & \\
                       &  (MAD = 80) & (MAD = 0.24) & (MAD = 0.03) & \\
\multirow{2}{*}{This Work -- Synthesis} & 32 $\pm$ 29 & 0.03 $\pm$ 0.05 & 0.05 $\pm$ 0.02 & 29 \\
%                       &  ($\sigma$ = 159) & ($\sigma$ = 0.26) & ($\sigma$ = 0.09) & \\
                       &  (MAD = 64) & (MAD = 0.15) & (MAD = 0.04) & \\
\multirow{2}{*}{This Work -- Photometry} & -12 $\pm$ 25 & 0.06 $\pm$ 0.02 & -- & 18 \\
%                       &  ($\sigma$ = 105) & ($\sigma$ = 0.08) & -- & \\
                       &  (MAD = 44) & (MAD = 0.04) & -- & \\
         \hline
\end{tabular}
}
\end{table}
%-------------------------------------------------------------------------

%----------------------------------------- FIGURE ---------------------------------------------------
%--------------------------------------------------------------------------------------------------

%----------------------------------------- FIGURE ---------------------------------------------------
%--------------------------------------------------------------------------------------------------

\section{Data and spectroscopic parameters for planet-hosts}\label{starsplanets}

We have identified spectra for 10 confirmed planet-hosts that show relatively high $\upsilon\sin i$ and we were unable 
to apply our standard EW method for their spectroscopic analysis. We use the procedure of this work to derive their stellar 
parameters to update the online SWEET-Cat catalogue where stellar parameters for FGK and M 
planet-hosts\footnote{https://www.astro.up.pt/resources/sweet-cat/} are presented \citep{santos13}. These stars were observed 
with high-resolution spectrographs (Table~\ref{table6}) gathered by our team (these spectra have never been analyzed before) 
and by the use of the archive (for the NARVAL spectra). Their spectral type varies from F to G.

The spectra were reduced with the standard pipelines and are corrected with the standard IRAF tools for the radial velocity 
shifts and in cases of multiple exposures of individual observed stars, their spectra are added. 
Following the procedure presented in this work, we derive their fundamental parameters, which are included in Figs.~\ref{fig:4} and 
\ref{fig:5} (square symbols) and presented in Table~\ref{table7}. The stellar masses and radii are calculated using the 
calibration of \cite{torres10_mass} with the corrections of \cite{santos13}.  

%----------------------------------------- FIGURE ---------------------------------------------------
\begin{figure}
  \centering
   \includegraphics[width=1.0\linewidth]{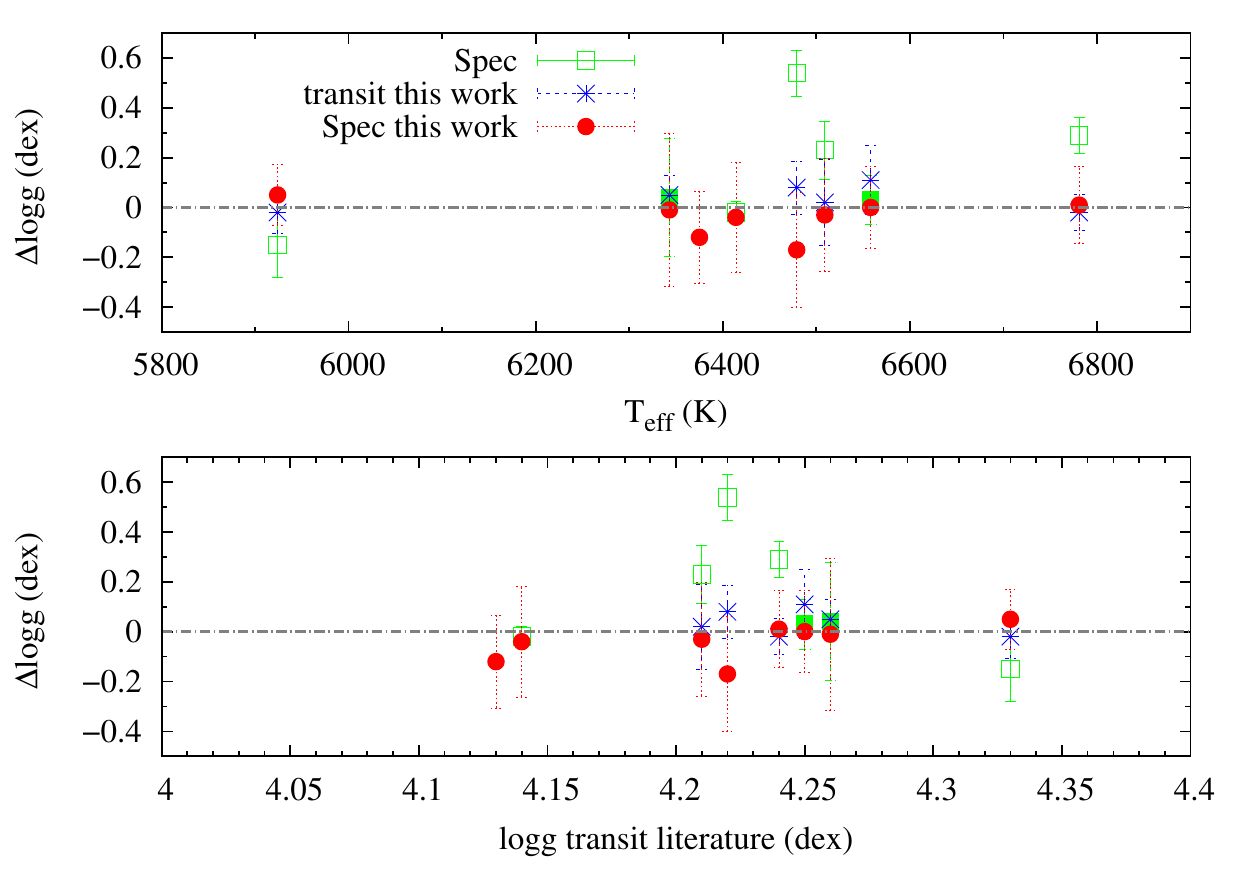}
  \caption{\footnotesize{The differences refer to surface gravity derived from a transit light curve analysis minus other methods from 
Table~\ref{table7}. Circles correspond to the comparison of $\log g$ derived in this work. Squares correspond to the methodology applied
by \cite{valenti05} and filled squares to other spectral synthesis methods. Asterisks show the comparison between $\log g$ derived from 
the light curve analysis on the literature and of this work.}}
  \label{fig:6}
  \end{figure}
%--------------------------------------------------------------------------------------------------

%----------------------------------Table --------------------------------------
\begin{table}
\begin{center}
\caption{Observation log of the transit hosts analyzed in this work. V magnitudes are taken from SIMBAD. S/N is estimated at 
6070 \AA{}. }
\label{table6}
\begin{tabular}{lcccc}
\hline\hline
Star & V (mag) & Spectrograph & Resolution & S/N  \\
\hline
30 Ari B & 7.09 & FEROS & 48000 & 270 \\
HAT-P-23 & 13.05 & FEROS & 48000 & 65 \\
HAT-P-34 & 10.40 & FEROS & 48000 & 145 \\
HAT-P-41 & 11.36 & FEROS & 48000 & 135 \\
HAT-P-2 & 8.69 & SOPHIE & 75000 & 250 \\
XO-3 & 9.85 & SOPHIE & 75000 & 130 \\
HD\,8673 & 6.31 & NARVAL & 75000 & 222 \\
Kepler-410A & 9.50 & NARVAL & 75000 & 72 \\
CoRoT-11 & 12.80 & HARPS & 110000 & 116 \\
CoRoT-3 & 13.29 & HARPS & 110000 & 84 \\
\hline
\end{tabular}
\end{center}
\end{table}
%--------------------------------------------------------------------------------------------------

%-----------------------------------------------------------------------
\begin{table*}
\begin{center}
\caption{Spectroscopic parameters of planet-hosts derived in this work and surface gravities derived from the transit light curve 
are found in the literature for all transiting planet-hosts of our sample.}
\label{table7}
\begin{tabular}{lcccccccccc}
\hline\hline
Star & $T_{\mathrm{eff}}$ & $\log g$ & $[Fe/H]$ & $\upsilon\sin i$ & $\log g_{transit} $ & Ref. & Mass & Radius \\
     & K & dex & dex & km/s & dex & & M$_{\odot}$ & R$_{\odot}$ \\
\hline
%HAT-P-23 fixed values   & 5924 $\pm$ 30 & 4.28 $\pm$ 0.11 & 0.16 $\pm$ 0.03 & This work & 5987 $\pm$ 120 & 4.48 $\pm$ 0.12 & 0.18 $\pm$ 0.10 & bakos & 8.50 $\pm$ 0.22 \\
HAT-P-23   & 5924 $\pm$ 30 & 4.28 $\pm$ 0.11 & 0.16  $\pm$ 0.03 & 8.50  $\pm$ 0.22  &  4.33 $\pm$ 0.05 & (1) & 1.13 $\pm$ 0.05 & 1.29 $\pm$ 0.05  \\
Kepler-410A & 6375 $\pm$ 44 & 4.25 $\pm$ 0.15 & 0.09  $\pm$ 0.04 & 13.24 $\pm$ 0.29 & 4.13 $\pm$ 0.11 & (2) & 1.30 $\pm$ 0.07 & 1.41 $\pm$ 0.07 \\
CoRoT-3    & 6558 $\pm$ 44 & 4.25 $\pm$ 0.15 & 0.14  $\pm$ 0.04 & 18.46 $\pm$ 0.29 & 4.25 $\pm$ 0.07 & (3) & 1.41 $\pm$ 0.08 & 1.44 $\pm$  0.08 \\
XO-3       & 6781 $\pm$ 44 & 4.23 $\pm$ 0.15 & -0.08 $\pm$ 0.04 & 18.77 $\pm$ 0.29 & 4.24 $\pm$ 0.04 & (4) & 1.41 $\pm$ 0.08 & 1.49 $\pm$ 0.08\\
HAT-P-41   & 6479 $\pm$ 51 & 4.39 $\pm$ 0.22 & 0.13  $\pm$ 0.05 & 20.11 $\pm$ 1.34 & 4.22 $\pm$ 0.07 & (5) & 1.28 $\pm$ 0.09 & 1.19 $\pm$ 0.08 \\
HAT-P-2    & 6414 $\pm$ 51 & 4.18 $\pm$ 0.22 & 0.04  $\pm$ 0.05 & 20.50 $\pm$ 1.34 & 4.14 $\pm$ 0.03 & (6) & 1.34 $\pm$ 0.09 & 1.54 $\pm$ 0.12 \\
HAT-P-34   & 6509 $\pm$ 51 & 4.24 $\pm$ 0.22 & 0.08  $\pm$ 0.05 & 24.08 $\pm$ 1.34 & 4.21 $\pm$ 0.06 & (7) & 1.36 $\pm$ 0.10 & 1.45 $\pm$ 0.11 \\
HD\,8673   & 6472 $\pm$ 51 & 4.27 $\pm$ 0.22 & 0.14  $\pm$ 0.05 & 26.91 $\pm$ 1.34 & -- & -- & 1.35 $\pm$ 0.10 & 1.39 $\pm$ 0.10 \\
CoRoT-11   & 6343 $\pm$ 72 & 4.27 $\pm$ 0.30 & 0.04  $\pm$ 0.03 & 36.72 $\pm$ 1.34 & 4.26 $\pm$ 0.06 & (8) & 1.56 $\pm$ 0.10 & 1.36 $\pm$ 0.13 \\
30\,Ari\,B & 6284 $\pm$ 98 & 4.35 $\pm$ 0.20 & 0.12  $\pm$ 0.08 & 42.61 $\pm$ 1.82 & -- & -- & 1.22 $\pm$ 0.08 & 1.23 $\pm$ 0.07 \\
\hline
\end{tabular}
\tablebib{(1)~\citet{bakos2011}; (2)~\citet{eylen2014}; (3)~\citet{deleuil2008}; (4)~\citet{johnskrull2008}; (5)~\citet{hartman2012}; 
(6)~\citet{pal2010}; (7)~\citet{bakos2012}; (8)~\citet{gandolfi2010}}
\end{center}
\end{table*}
%I used the spec logg but the constrained other parameters
%only corot-3 and 11 are derived with mean of other methods
%------------------------------------------------------------------------------------------------------------------
%------------------------------------------------------------------------------------------------------------------
\begin{table}
\centering
\caption{Transit fit parameters}
\label{tab:transit_fit}
\begin{tabular}{lcccc}
\hline\hline
Name & R$_p$/R$_{\star}$ & T$_{d} $ & $\rho_{\star}$ & g$_1$ \\
     &                   & days          & g cm$^{-3}$    &      \\
\hline
HAT-P-23 & 0.1209$_{-0.0011}^{+0.0015}$ & 0.0822$_{-0.0008}^{+0.0005}$ & 0.976$_{-0.102}^{+0.068}$ & 0.281$_{-0.056}^{+0.037}$ \\
HAT-P-34 & 0.0842$_{-0.0015}^{+0.0015}$ & 0.1323$_{-0.0015}^{+0.0013}$ & 0.505$_{-0.119}^{+0.097}$ & 0.037$_{-0.019}^{+0.111}$ \\
HAT-P-41 & 0.1049$_{-0.0004}^{+0.0011}$ & 0.1523$_{-0.0009}^{+0.0004}$ & 0.452$_{-0.054}^{+0.003}$ & 0.211$_{-0.044}^{+0.019}$ \\
XO-3     & 0.0915$_{-0.0007}^{+0.0006}$ & 0.1043$_{-0.0008}^{+0.0010}$ & 0.649$_{-0.060}^{+0.060}$ & 0.343$_{-0.090}^{+0.030}$ \\
CoRoT-3  & 0.0641$_{-0.0005}^{+0.0007}$ & 0.1410$_{-0.0008}^{+0.0010}$ & 0.431$_{-0.055}^{+0.074}$ & 0.202$_{-0.062}^{+0.041}$ \\
CoRoT-11 & 0.0999$_{-0.0005}^{+0.0006}$ & 0.0799$_{-0.0012}^{+0.0010}$ & 0.581$_{-0.023}^{+0.034}$ & 0.347$_{-0.092}^{+0.062}$ \\
\hline
\end{tabular}
\end{table}
%------------------------------------------------------------------------------------------------------------------

\subsection{Transit analysis}

We retrieve from the literature available photometric data for our transiting planet target stars. Our aim is to perform an 
homogeneous analysis of these objects using our re-determined stellar parameters to guess limb darkening coefficients and average 
stellar density. The limb darkening coefficients are linearly interpolated in the 4 dimension of the new stellar parameters 
($T_{\mathrm{eff}}$, $\log g$,  $[Fe/H]$, and $\upsilon_{mac}$) from the tables of \cite{claret11} to match our stellar parameter values.
We obtain as well the stellar density from the mass and radius as described in the previous Section. Transit duration and 
transit depth are initially taken from the values quoted in the literature. The light curves are all folded with the period known
from the literature and out of transit measurements are normalized to one.

Since some of the planets in our sample are in eccentric orbits, we adopt the expansion to the fourth order for the normalized 
projected distance of the planet with respect to the stellar center reported in \cite{pal2010} and express it as a function of 
the stellar density ($\rho_{\star}$) and the transit duration ($T_d$).

For each folded light curve, we fit a transiting planet model using the \cite{mandel02} model and the Levemberg-Marquardt algorithm 
\citep{press}. For eccentric planets we adopt the values of the eccentricity and argument of periastron reported in the literature
and add a gaussian prior on both during our error analysis (see below) considering the reported uncertainties.

The uncertainties of the measurements are first expanded by the reduced $\chi^{2}$ of the fit. We account for correlated noise 
creating a  mock  sample of the fit residuals (using the  measurement uncertainties) and comparing the scatter in the artificial and 
in the real light curves re-binning the residuals on increasing time-intervals (up to 30 min). If the ratio of the expected to the 
real scatter is found larger than one, we further expanded the uncertainties by this factor. Finally, we determine the distributions 
of the parameter best-fit values bootstrapping the  light curves and derived the mode of the resulting distributions, and the 68.3 
per cent confidence limits defined by the 15.85th and the 84.15th percentiles in the cumulative distributions.

The results are reported in Table~\ref{tab:transit_fit}. The photometric densities appear smaller than the values implied by 
theoretical models. The discrepancy is largest for the case of Kepler-410A where models predict $\rho_{\star}\sim1\,$g  cm$^{-3}$, 
whereas the measured value is 0.0937$_{-0.0052}^{+0.0070}$ g cm$^{-3}$. The dilution caused by the contamination of a stellar companion 
(Kepler-410B) and the small size of the planet (2.838\,R$_{\oplus}$, \citet{eylen2014}) are the main reasons for the difference in 
the density derived from the transit fit. Considering the above, we exclude this star from the comparison of the transit fit results.

\subsection{Discussion}

For stars with a transiting planet, surface gravity has been proposed to be independently derived from the light curve with better 
precision than from spectroscopy \citep{seager2003, sozzetti2007}. 
In \cite{torres12}, it has been shown that $\log g$ derived using SME and the methodology of \cite{valenti05} is 
systematically underestimated for hotter stars ($T_{\mathrm{eff}}>$\,6000\,K) when compared with the $\log g$ from transit fits. 
According to the authors, constraining $\log g$ to the transit values, as more reliable, leads to significant biases in the temperature 
and metallicitity which consequently propagates to biases in stellar (and planetary) mass and radius. 

From the planet-hosts in our work, there are 8 stars with transit data and available $\log g$ using a light curve analysis. We therefore, 
compare the $\log g$ derived from our spectroscopic analysis with the $\log g$ from the transit fits as taken from the literature 
(red circles in Fig.~\ref{fig:6}). The differences of this comparison are very small ($\Delta \log g$ =\,-0.04\, with 
$\sigma$=\,0.07\,dex). On the other hand, a comparison between the $\log g$ from the transit light curve and the $\log g$ using only the 
unconstrained \cite{valenti05} methodology shows difference of 0.18\,($\sigma$=\,0.27)\,dex for 5 stars with available measurements 
(empty squares in Fig.~\ref{fig:6}). We also plot for completeness the $\log g$ from our light curve analysis of the previous Section, 
using the stellar density and mass (asterisks in Fig.~\ref{fig:6}). 

Even though the number of stars for this comparison is very small, these results suggest that fixing $\log g$ to the transit value is not 
required with the analysis of this work, avoiding the biases that are described in \cite{torres12}. The different approach we adopt in 
this work, mainly due to the different line list, shows that we obtain a better estimate on surface gravity. However, since our sample is 
small and limited only to hotter stars, further investigation is advised to check whether following the unconstrained approach is the 
optimal strategy. The unconstrained analysis is also suggested in \cite{gomez2013} as preferred, after analyzing the transit-host WASP-13 
with SME but following different methodology (line list, initial parameters, convergence criteria, fixed parameters) from \cite{valenti05}. 
%In Fig.~\ref{fig:6} we show the comparison of surface gravities derived from the transit light curve with the $\log g$ derived 
%spectroscopically. 

%----------------------------------------- FIGURE ---------------------------------------------------
\begin{figure}
  \centering
   \includegraphics[width=1.0\linewidth, height =0.4\textheight]{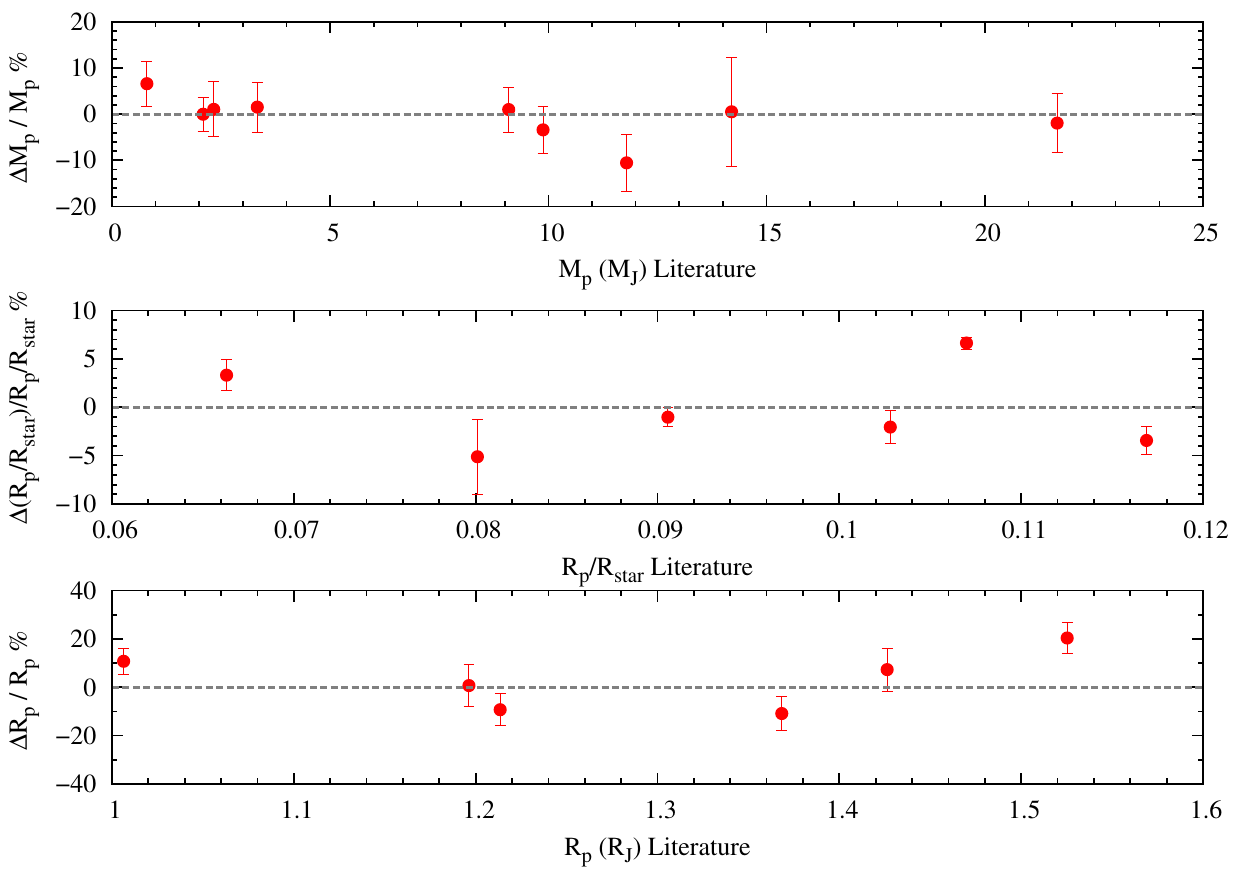}
  \caption{\footnotesize{Comparison between the literature data of planetary mass, the radii ratio (R$_{p}$/R$_{star}$), and planetary 
radius and this work, respectively in absolute units.}}
  \label{planet_values}
  \end{figure}
%--------------------------------------------------------------------------------------------------

We explore how the literature values of planetary mass and radius are affected with the new stellar parameters. From our analysis we find 
that the dispersion between the planetary mass derived with our stellar parameters and the literature is 4\%  
(Fig.~\ref{planet_values}, top panel). The planet-to-star radius ratio derived from the transit light curve shows 
same dispersion of 4\% (Fig.~\ref{planet_values}, middle panel). This consistency with the literature values confirms the accuracy of the 
transit light curve analysis to derive planet-to-star radius ratio. The planetary radius is calculated from this ratio and the stellar 
radius that is inferred from our spectroscopic values. The comparison of the planetary radius with the literature values shows the 
highest dispersion of 14\% (Fig.~\ref{planet_values}, bottom panel). Since we have shown the consistency of the 
planet-to-star radius ratio, the main source of uncertainty in the derivation of planetary radius is the calculation of the stellar value. 

%----------------------------------------- FIGURE ---------------------------------------------------
\begin{figure}
  \centering
   \includegraphics[width=1.0\linewidth]{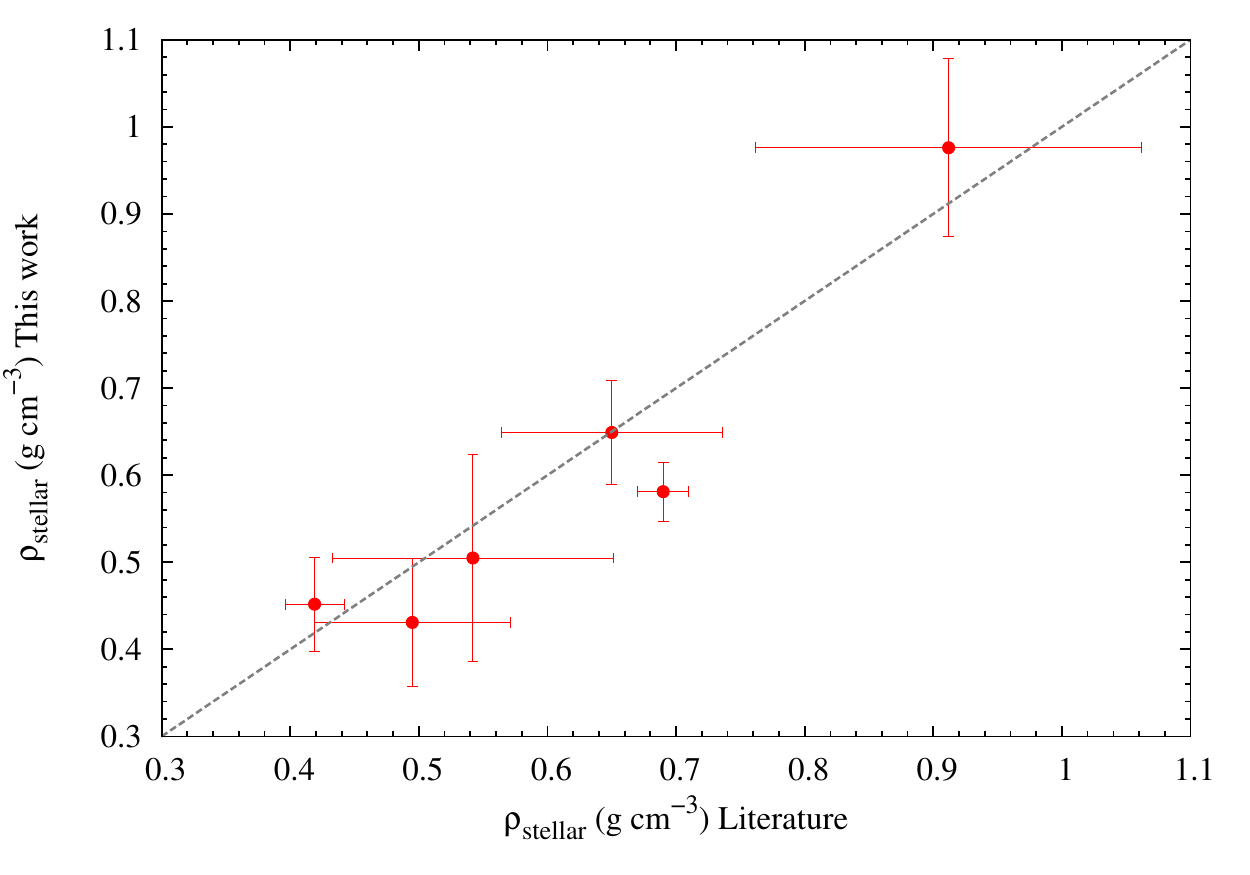}
  \caption{\footnotesize{Comparison between stellar density derived from the transit light curve analysis and literature data.}}
  \label{density}
  \end{figure}
%--------------------------------------------------------------------------------------------------

%----------------------------------------- FIGURE ---------------------------------------------------
\begin{figure}
  \centering
   \includegraphics[width=1.0\linewidth]{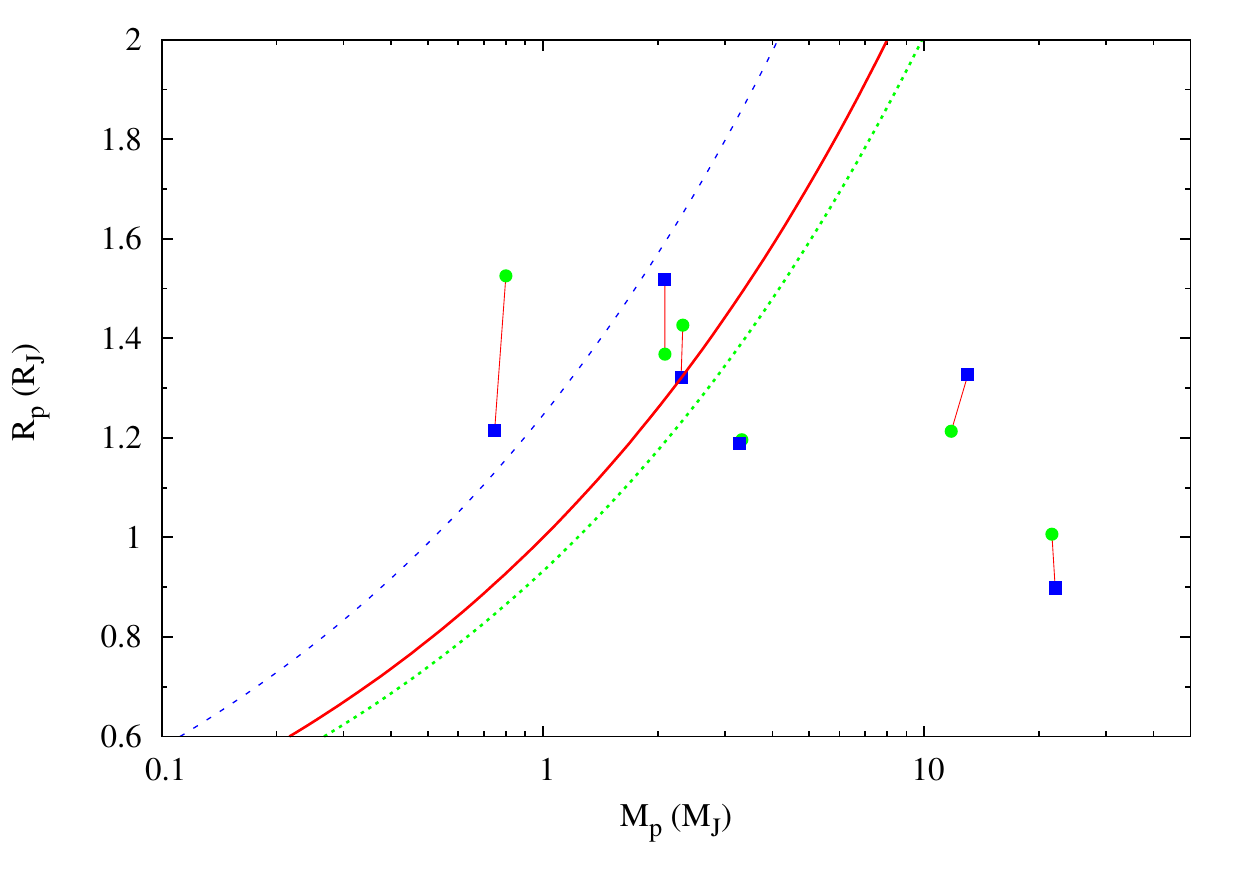}
  \caption{\footnotesize{Blue squares represent planetary mass and radius derived in this work in comparison with literature values 
(green circles). Characteristic isodensity curves are plotted for Saturn (dashed), Jupiter (solid) and Neptune (dotted).}}
  \label{mass_radius}
  \end{figure}
%--------------------------------------------------------------------------------------------------

We also compare the stellar density derived from the transit analysis with the respective ones from the literature (Fig.~\ref{density}). 
In Fig.~\ref{mass_radius}, we show the new mass and radius from this work in comparison with the literature values. Planetary radius 
shows higher discrepancies mainly because of the uncertainties in the stellar radius calculations. 

\section{Conclusions}

In this paper we are introducing a new approach for the derivation of the fundamental stellar parameters for FGK dwarfs using the 
spectral synthesis technique. In particular, we focus on stars with moderate/high rotational velocities. Such stars could be transiting 
planet-hosts as they show high dispersion in their rotational velocities and the determination of their stellar parameters is of 
principal importance for the planetary studies. 

The key of our method is the selection of the line list that contains principally iron lines based on previous work. This 
line list is tested primarily for stars with low rotational velocities. The comparison in temperatures between the EW method and this work 
for our test sample is in good agreement even though high temperatures ($T_{\mathrm{eff}}>$ 6000\,K) are underestimated by the synthesis 
technique. Metallicity is in excellent agreement with our test sample and surface gravity shows an offset of -0.19\,dex.

Our method is applied to reference stars that are convolved with a set of rotational profiles (up to 
$\upsilon\sin i$\,=\,50\,km/s). The spectrum of each reference star has been broadened with different $\upsilon\sin i$ values and for 
these spectra we calculate the stellar parameters which are compared with the initial unbroadened spectrum in order to check their 
consistency for high $\upsilon\sin i$. The results show that even for high $\upsilon\sin i$ the differences from the ones without 
broadening is on the same scale as the errors. 

As a final test we calculate stellar parameters for a sample of stars with high $\upsilon\sin i$ and compare with other spectral 
synthesis methodologies, the EW method (when possible), and the IRFM. The comparison shows very good agreement with all methods with the 
same dispersion in the mean differences of the parameters as the slow rotating stars.  

We also applied our method to 10 planet-hosts with moderate/high rotation, updating their stellar parameters in a uniform way. 
A new analysis has been conducted to the light curves for the stars that had available photometric observations using our results. 
From the combination of spectroscopic parameters and the ones derived from the transit fits (namely stellar density), we calculate 
surface gravity and the comparison with spectroscopic derivations suggests that fixing $\log g$ to the transit value is not required 
using our method. In addition, we present the difference in the planetary mass and radius (expressed as a percentage) with the literature 
values. Planetary masses agree very well with the literature values. The dispersion in the radii of the planets is higher due 
to larger errors in the estimation of the stellar radius. 

The study of planet-hosts with higher rotational velocities is essential because they expand the planet sample around stars of earlier 
types (F- and A-type) that are more massive than the Sun. Precise stellar parameters for these stars are necessary to study the frequency 
of planets around intermediate mass stars and explore their planet formation mechanisms. Additionally, precise (and if possible accurate) 
stellar parameters are essential for a detailed characterization of the planets to be discovered by the upcoming high precision transit 
missions such as CHEOPS, TESS, and PLATO\,2.0.

\begin{acknowledgements}
This work is supported by the European Research 
Council/European Community under the FP7 through Starting Grant agreement number 239953. 
N.C.S. also acknowledges the support from Funda\c{c}\~ao para a Ci\^encia e a Tecnologia (FCT) through program Ci\^encia 2007 funded 
by FCT/MCTES (Portugal) and POPH/FSE (EC), and in the form of grant reference PTDC/CTE-AST/098528/2008. 
S.G.S, E.D.M, and V.Zh.A. acknowledge the support from the Funda\c{c}\~ao para a
Ci\^encia e Tecnologia, FCT (Portugal) in the form of the fellowships
 SFRH/BPD/47611/2008, SFRH/BPD/76606/2011, and SFRH/BPD/70574/2010. 
G.I. acknowledges financial support from the Spanish Ministry project MINECO AYA2011-29060.
This research has made use of the SIMBAD database operated at CDS, Strasbourg, France and the Vienna 
Atomic Line Database operated at Uppsala University, the Institute of Astronomy RAS in Moscow, and the University of Vienna.
We thank the authors of SME for making their code public. The Narval observations were conducted under the OPTICON access programme.
OPTICON has received research funding from the European Community's Sixth Framework Programme under contract number RII3-CT-001566.
\end{acknowledgements}

\bibliography{bibliography}

\begin{appendix}
\section{}
%-----------------------------------------------------------------------
  \begin{table*}
     \centering
     \caption[]{Results of the comparison between this work and the EW method. The stars in boldface are analyzed in 
Sect.~\ref{rotation}.}
     \label{table8}
        \begin{tabular}{lccccccc}
           \hline\hline
     &    \multicolumn{4}{c}{This work}   & \multicolumn{3}{c}{EW method} \\
      \cline{1-5}  \cline{6-8}
Star & $T_{\mathrm{eff}}$ & $\log g$ & [Fe/H] &  $\upsilon\sin i$ & $T_{\mathrm{eff}}$ & $\log g$ & [Fe/H] \\
           & (K) & (dex) & (dex) & (km/s) & (K) & (dex) & (dex) \\
\hline
\multicolumn{8}{c}{Dwarf stars} \\
\hline
CoRoT-2     & 5620 $\pm$ 18 & 4.66 $\pm$ 0.06 & -0.03 $\pm$ 0.03 & 9.97 & 5697 $\pm$ 97 & 4.73 $\pm$ 0.17 & -0.09 $\pm$ 0.07 \\
CoRoT-10    & 4921 $\pm$ 25 & 4.09 $\pm$ 0.09 & 0.15 $\pm$ 0.03 & 2.19 & 5025 $\pm$ 155 & 4.47 $\pm$ 0.31 & 0.06 $\pm$ 0.09 \\
CoRoT-4     & 6164 $\pm$ 30 & 4.34 $\pm$ 0.11 & 0.15 $\pm$ 0.03 & 7.03 & 6344 $\pm$ 93 & 4.82 $\pm$ 0.11 & 0.15 $\pm$ 0.06 \\
CoRoT-5     & 6254 $\pm$ 30 & 4.41 $\pm$ 0.11 & 0.04 $\pm$ 0.03 & 1.43 & 6240 $\pm$ 70 & 4.46 $\pm$ 0.11 & 0.04 $\pm$ 0.05 \\
HD\,101930  & 5083 $\pm$ 18 & 4.15 $\pm$ 0.06 & 0.10 $\pm$ 0.03 & 0.10 & 5083 $\pm$ 63 & 4.35 $\pm$ 0.13 & 0.16 $\pm$ 0.04 \\
HD\,102365  & 5588 $\pm$ 18 & 4.07 $\pm$ 0.06 & -0.30 $\pm$ 0.03 & 0.10 & 5616 $\pm$ 41 & 4.40 $\pm$ 0.06 & -0.28 $\pm$ 0.03 \\
HD\,103774  & 6582 $\pm$ 30 & 4.47 $\pm$ 0.11 & 0.27 $\pm$ 0.03 & 8.93 & 6732 $\pm$ 56 & 4.81 $\pm$ 0.06 & 0.29 $\pm$ 0.03 \\
HD\,1237    & 5588 $\pm$ 18 & 4.58 $\pm$ 0.06 & 0.11 $\pm$ 0.03 & 4.62 & 5489 $\pm$ 40 & 4.46 $\pm$ 0.11 & 0.06 $\pm$ 0.03 \\
HD\,134060  & 5914 $\pm$ 18 & 4.28 $\pm$ 0.06 & 0.09 $\pm$ 0.03 & 1.44 & 5940 $\pm$ 18 & 4.42 $\pm$ 0.03 & 0.12 $\pm$ 0.01 \\
HD\,1388    & 5967 $\pm$ 18 & 4.38 $\pm$ 0.06 & 0.00 $\pm$ 0.03 & 1.27 & 5970 $\pm$ 15 & 4.42 $\pm$ 0.05 & 0.00 $\pm$ 0.01 \\
HD\,148156  & 6212 $\pm$ 30 & 4.40 $\pm$ 0.11 & 0.23 $\pm$ 0.03 & 5.73 & 6251 $\pm$ 25 & 4.51 $\pm$ 0.05 & 0.25 $\pm$ 0.02 \\
HD\,162020  & 4798 $\pm$ 25 & 4.14 $\pm$ 0.09 & -0.14 $\pm$ 0.03 & 1.46 & 4723 $\pm$ 71 & 4.31 $\pm$ 0.18 & -0.10 $\pm$ 0.03 \\
\bf{HD\,20852} & 6675 $\pm$ 30 & 4.12 $\pm$ 0.11 & -0.37 $\pm$ 0.03 & 7.06 & 6813 $\pm$ 92 & 4.76 $\pm$ 0.12 & -0.35 $\pm$ 0.06 \\
\bf{HD\,20868} & 4745 $\pm$ 25 & 4.02 $\pm$ 0.09 & 0.00 $\pm$ 0.03 & 0.46 & 4720 $\pm$ 91 & 4.24 $\pm$ 0.47 & 0.08 $\pm$ 0.01 \\
HD\,221287  & 6337 $\pm$ 30 & 4.43 $\pm$ 0.06 & 0.02 $\pm$ 0.06 & 3.92 & 6417 $\pm$ 25 & 4.60 $\pm$ 0.10 & 0.06 $\pm$ 0.02 \\
HD\,222237  & 4618 $\pm$ 25 & 3.92 $\pm$ 0.09 & -0.50 $\pm$ 0.03 & 0.10 & 4722 $\pm$ 55 & 4.34 $\pm$ 0.15 & -0.39 $\pm$ 0.06 \\
HD\,23079   & 5965 $\pm$ 18 & 4.28 $\pm$ 0.06 & -0.13 $\pm$ 0.03 & 0.10 & 6009 $\pm$ 14 & 4.50 $\pm$ 0.05 & -0.11 $\pm$ 0.01 \\
\bf{HD\,27894} & 4894 $\pm$ 25 & 4.08 $\pm$ 0.09 & 0.18 $\pm$ 0.03 & 0.87 & 4833 $\pm$ 209 & 4.30 $\pm$ 0.48 & 0.26 $\pm$ 0.10 \\
HD\,31527   & 5915 $\pm$ 18 & 4.40 $\pm$ 0.06 & -0.17 $\pm$ 0.03 & 2.36 & 5917 $\pm$ 13 & 4.47 $\pm$ 0.05 & -0.17 $\pm$ 0.01 \\
HD\,330075  & 4924 $\pm$ 30 & 4.03 $\pm$ 0.09 & -0.04 $\pm$ 0.03 & 0.10 & 4958 $\pm$ 52 & 4.24 $\pm$ 0.13 & 0.05 $\pm$ 0.03 \\
HD\,361     & 5924 $\pm$ 18 & 4.48 $\pm$ 0.06 & -0.10 $\pm$ 0.03 & 0.10 & 5888 $\pm$ 14 & 4.54 $\pm$ 0.08 & -0.13 $\pm$ 0.01 \\
HD\,38283   & 5962 $\pm$ 18 & 4.14 $\pm$ 0.06 & -0.15 $\pm$ 0.03 & 4.51 & 5980 $\pm$ 24 & 4.27 $\pm$ 0.03 & -0.14 $\pm$ 0.02 \\
\bf{HD\,40307} & 4771 $\pm$ 25 & 4.10 $\pm$ 0.09 & -0.42 $\pm$ 0.03 & 0.10 & 4774 $\pm$ 77 & 4.42 $\pm$ 0.16 & -0.36 $\pm$ 0.02 \\
\bf{HD\,61421} & 6616 $\pm$ 30 & 4.09 $\pm$ 0.11 & 0.03 $\pm$ 0.03 & 4.40 & 6612 & 4.02 & -0.02 \\
\bf{HD\,63454} & 4833 $\pm$ 25 & 4.11 $\pm$ 0.09 & 0.04 $\pm$ 0.03 & 1.81 & 4756 $\pm$ 77 & 4.32 $\pm$ 0.22 & 0.13 $\pm$ 0.05 \\
HD\,750     & 5118 $\pm$ 18 & 4.34 $\pm$ 0.06 & -0.29 $\pm$ 0.03 & 0.10 & 5069 $\pm$ 32 & 4.33 $\pm$ 0.1 & -0.30 $\pm$ 0.02 \\
HD\,870     & 5379 $\pm$ 18 & 4.36 $\pm$ 0.06 & -0.12 $\pm$ 0.03 & 0.10 & 5360 $\pm$ 24 & 4.40 $\pm$ 0.08 & -0.12 $\pm$ 0.02 \\
HD\,93385   & 5987 $\pm$ 18 & 4.38 $\pm$ 0.06 & 0.02 $\pm$ 0.03 & 1.06 & 5989 $\pm$ 17 & 4.46 $\pm$ 0.03 & 0.03 $\pm$ 0.01 \\
HD\,967     & 5643 $\pm$ 18 & 4.38 $\pm$ 0.06 & -0.59 $\pm$ 0.03 & 0.10 & 5595 $\pm$ 18 & 4.59 $\pm$ 0.02 & -0.66 $\pm$ 0.01 \\
OGLE-TR-113 & 4793 $\pm$ 25 & 4.25 $\pm$ 0.09 & 0.05 $\pm$ 0.03 & 5.02 & 4781 $\pm$ 166 & 4.31 $\pm$ 0.41 & 0.03 $\pm$ 0.06 \\
WASP-29     & 4782 $\pm$ 25 & 4.13 $\pm$ 0.09 & 0.18 $\pm$ 0.03 & 0.10 & 5203 $\pm$ 102 & 4.93 $\pm$ 0.21 & 0.17 $\pm$ 0.05 \\
%WASP-17     & 6666 $\pm$ 30 & 4.26 $\pm$ 0.11 & -0.04 $\pm$ 0.03 & 9.93 & 6794 $\pm$ 83 & 4.83 $\pm$ 0.09 & -0.12 $\pm$ 0.05 \\
WASP-15     & 6378 $\pm$ 30 & 4.24 $\pm$ 0.11 & 0.03 $\pm$ 0.03 & 5.13 & 6573 $\pm$ 70 & 4.79 $\pm$ 0.08 & 0.09 $\pm$ 0.03 \\
WASP-16     & 5710 $\pm$ 18 & 4.23 $\pm$ 0.06 & 0.12 $\pm$ 0.03 & 0.47 & 5726 $\pm$ 22 & 4.34 $\pm$ 0.05 & 0.13 $\pm$ 0.02 \\
WASP-17     & 6666 $\pm$ 30 & 4.26 $\pm$ 0.06 & -0.04 $\pm$ 0.03 & 9.93 & 6794 $\pm$ 83 & 4.83 $\pm$ 0.09 & -0.12 $\pm$ 0.05 \\
WASP-2      & 5105 $\pm$ 18 & 3.97 $\pm$ 0.06 & 0.08 $\pm$ 0.03 & 2.90 & 5109 $\pm$ 72 & 4.33 $\pm$ 0.14 & 0.02 $\pm$ 0.05 \\
WASP-23     & 5053 $\pm$ 18 & 4.20 $\pm$ 0.06 & -0.02 $\pm$ 0.03 & 0.46 & 5046 $\pm$ 99 & 4.33 $\pm$ 0.18 & 0.05 $\pm$ 0.06 \\
WASP-38     & 6247 $\pm$ 30 & 4.25 $\pm$ 0.11 & 0.06 $\pm$ 0.03 & 8.05 & 6436 $\pm$ 60 & 4.80 $\pm$ 0.07 & 0.06 $\pm$ 0.04 \\
WASP-6      & 5447 $\pm$ 18 & 4.42 $\pm$ 0.06 & -0.11 $\pm$ 0.03 & 0.10 & 5383 $\pm$ 41 & 4.52 $\pm$ 0.06 & -0.14 $\pm$ 0.03 \\
\bf{Sun}    & 5771 $\pm$ 18 & 4.42 $\pm$ 0.06 & 0.00 $\pm$ 0.03 & 2.57 & -- & -- & -- \\ 
         \hline
 \multicolumn{8}{c}{Giant stars} \\
  \hline
HD\,148427    & 5018 $\pm$ 25 & 3.49 $\pm$ 0.09 & 0.01  $\pm$ 0.03 & 0.45 & 4962 $\pm$ 45 & 3.39 $\pm$ 0.12 & 0.03  $\pm$ 0.03 \\
HD\,175541    & 5097 $\pm$ 18 & 3.44 $\pm$ 0.06 & -0.14 $\pm$ 0.03 & 2.45 & 5111 $\pm$ 38 & 3.56 $\pm$ 0.08 & -0.11 $\pm$ 0.03 \\
HD\,27442     & 4852 $\pm$ 25 & 3.48 $\pm$ 0.09 & 0.23  $\pm$ 0.03 & 2.65 & 4781 $\pm$ 76 & 3.46 $\pm$ 0.19 & 0.33  $\pm$ 0.05 \\
HD\,62509     & 5007 $\pm$ 25 & 3.06 $\pm$ 0.09 & 0.21  $\pm$ 0.03 & 3.76 & 4935 $\pm$ 49 & 2.91 $\pm$ 0.13 & 0.09  $\pm$ 0.04 \\
HD\,88133     & 5330 $\pm$ 18 & 3.62 $\pm$ 0.06 & 0.20  $\pm$ 0.03 & 3.39 & 5438 $\pm$ 34 & 3.94 $\pm$ 0.11 & 0.33  $\pm$ 0.05 \\
HD\,142091    & 4898 $\pm$ 25 & 3.24 $\pm$ 0.09 & 0.05  $\pm$ 0.03 & 4.38 & 4876 $\pm$ 46 & 3.15 $\pm$ 0.14 & 0.13  $\pm$ 0.03 \\ 
HD\,188310    & 4799 $\pm$ 18 & 3.14 $\pm$ 0.06 & -0.06 $\pm$ 0.03 & 5.28 & 4714 $\pm$ 49 & 2.53 $\pm$ 0.11 & -0.27 $\pm$ 0.04 \\
HD\,163917    & 5107 $\pm$ 18 & 2.82 $\pm$ 0.06 & 0.33  $\pm$ 0.03 & 4.21 & 4967 $\pm$ 61 & 2.70 $\pm$ 0.13 & 0.14  $\pm$ 0.05 \\
\hline
        \end{tabular}
  \end{table*}
%-------------------------------------------------------------------------

\begin{landscape}
\begin{table}
%\centering
\caption{Stellar parameters for a sample of high rotating FGK dwarfs.}
\label{table9}
%\rotatebox{90}{
\scalebox{0.86}{
\begin{tabular}{lcccccccccccccccccccccccccccccc}
\hline\hline
Star & $T_{\mathrm{eff} IRFM}$ & Ref. & $\log g_{HIP}$ & $T_{\mathrm{eff}}$ & $\log g$ & $[Fe/H]$ & Ref. & $T_{\mathrm{eff} Synth}$ & $\log g_{Synth}$ & $[Fe/H]_{Synth}$ & Ref. & $T_{\mathrm{eff} EW}$ & $\log g_{EW}$ & $[Fe/H]_{EW}$ & Ref. & $\upsilon\sin i$ \\
     & K &  & dex & K & dex & dex &  & K & dex & dex &  & K & dex & dex &  & km/s \\
\hline 
HD\,179949  & 6205 $\pm$ 80  & (1)  & 4.38 $\pm$ 0.10 & 6237 $\pm$ 30 & 4.40 $\pm$ 0.11 & 0.17  $\pm$ 0.03 & This work & 6168 $\pm$ 44  & 4.34 $\pm$ 0.06 & 0.11  $\pm$ 0.03 & (3)  & 6287 $\pm$ 28 & 4.54 $\pm$ 0.04 & 0.21 $\pm$ 0.02 & 18 & 6.52 \\
HD\,165185  & 5932 $\pm$ 80  & (1)  & 4.47 $\pm$ 0.10 & 5940 $\pm$ 18 & 4.46 $\pm$ 0.06 & -0.05 $\pm$ 0.03 & This work & 5906 & 4.44 & -0.07 & (4) & 5942 $\pm$ 85 & 4.53 $\pm$ 0.13 & 0.02 $\pm$ 0.10 & 19 & 7.53 \\
HAT-P-6     & --             & -- & --                & 6933 $\pm$ 30 & 4.38 $\pm$ 0.11 & -0.02 $\pm$ 0.03 & This work & 6353 $\pm$ 88  & 3.84 $\pm$ 0.12 & -0.23 $\pm$ 0.08 & (5)  & 6855 $\pm$ 111 & 4.69 $\pm$ 0.20 & -0.08 $\pm$ 0.11 & 20 & 8.06 \\
HAT-P-23    & --	     & -- & --                & 5924 $\pm$ 30 & 4.28 $\pm$ 0.11 & 0.16  $\pm$ 0.03 & This work & 5905 $\pm$ 80  & 4.48 $\pm$ 0.12 & 0.15 $\pm$ 0.04 & (10) & -- & -- & -- & -- & 8.50 \\
HD\,19994   & 6159 $\pm$ 80  & (1)  & 4.10 $\pm$ 0.10 & 6145 $\pm$ 30 & 4.10 $\pm$ 0.11 & 0.20  $\pm$ 0.03 & This work & 6188 $\pm$ 44  & 4.24 $\pm$ 0.06 & 0.17  $\pm$ 0.03 & (3)  & 6289 $\pm$ 46 & 4.48 $\pm$ 0.05 & 0.24 $\pm$ 0.03 & 18 & 8.51 \\
HD\,89744   & 6262 $\pm$ 92  & (1)  & 3.97 $\pm$ 0.10 & 6300 $\pm$ 30 & 4.07 $\pm$ 0.11 & 0.24  $\pm$ 0.03 & This work & 6291 $\pm$ 44  & 4.07 $\pm$ 0.06 & 0.20  $\pm$ 0.03 & (3)  & 6234 $\pm$ 45 & 3.98 $\pm$ 0.05 & 0.22 $\pm$ 0.05 & 21 & 8.86 \\
HD\,49933   & 6609 $\pm$ 80  & (1)  & 4.40 $\pm$ 0.10 & 6904 $\pm$ 44 & 4.18 $\pm$ 0.15 & -0.17 $\pm$ 0.04 & This work & 6780 $\pm$ 70  & 4.30 $\pm$ 0.20 & -0.30 $\pm$ 0.11 & (6)  & 6522  & 4.00 & -0.49 & 22 & 10.14 \\
HD\,142     & 6313 $\pm$ 80  & (1)  & 4.27 $\pm$ 0.10 & 6271 $\pm$ 44 & 4.17 $\pm$ 0.15 & 0.13  $\pm$ 0.04 & This work & 6249 $\pm$ 44  & 4.19 $\pm$ 0.06 & 0.08  $\pm$ 0.03 & (3)  & 6403 $\pm$ 65 & 4.62 $\pm$ 0.07 & 0.09 $\pm$ 0.05 & 18 & 10.22 \\
HD\,142860  & 6336 $\pm$ 80  & (1)  & 4.27 $\pm$ 0.10 & 6361 $\pm$ 44 & 4.07 $\pm$ 0.15 & -0.09 $\pm$ 0.04 & This work & 6262 $\pm$ 44  & 4.18 $\pm$ 0.06 & -0.14 $\pm$ 0.03 & (3)  & 6281 & 4.06 & -0.13 & 22 & 10.65 \\
HD\,89569   & 6439 $\pm$ 80  & (1)  & 4.12 $\pm$ 0.10 & 6469 $\pm$ 44 & 4.08 $\pm$ 0.15 & 0.09  $\pm$ 0.04 & This work & 6401 & 3.99 & -0.12 & (4)  & -- & -- & -- & --  & 11.33 \\
HD\,86264   & 6381 $\pm$ 80  & (1)  & 4.16 $\pm$ 0.10 & 6300 $\pm$ 44 & 4.06 $\pm$ 0.15 & 0.25  $\pm$ 0.04 & This work & 6326 $\pm$ 44  & 4.22 $\pm$ 0.05 & 0.16  $\pm$ 0.03 & (3)  & 6596 $\pm$ 78 & 4.47 $\pm$ 0.15 & 0.37 $\pm$ 0.06 & 23 & 12.55 \\
HD\,121370  & 6141 $\pm$ 80  & (1)  & 3.83 $\pm$ 0.10 & 6080 $\pm$ 44 & 3.78 $\pm$ 0.15 & 0.33  $\pm$ 0.04 & This work & 6030 $\pm$ 80  & 3.90 $\pm$ 0.08 & 0.24  $\pm$ 0.07 & (7)  & 6300 &  4.18 & 0.29 & 22 & 13.10 \\
Kepler-410A & 6273 $\pm$ 140 & (2)  & --              & 6375 $\pm$ 44 & 4.25 $\pm$ 0.15 & 0.09  $\pm$ 0.04 & This work & 6325 $\pm$ 75  & -- & 0.01 $\pm$ 0.10 & 8 & -- & -- & -- & --  &  13.24 \\
HD\,210302  & 6477 $\pm$ 80  & (1)  & 4.29 $\pm$ 0.10 & 6405 $\pm$ 44 & 4.24 $\pm$ 0.15 & 0.10  $\pm$ 0.04 & This work & 6339 $\pm$ 44  & 4.15 $\pm$ 0.06 & 0.08  $\pm$ 0.03 & (3)  & -- & -- & -- & --  & 13.68 \\
HD\,105     & 6035 $\pm$ 80  & (1)  & 4.46 $\pm$ 0.10 & 6045 $\pm$ 44 & 4.40 $\pm$ 0.15 & 0.02  $\pm$ 0.04 & This work & 6126 $\pm$ 44  & 4.65 $\pm$ 0.06 & -0.02 $\pm$ 0.03 & (3)  & 6012 $\pm$ 68  & 4.42 $\pm$ 0.12 & 0.06 $\pm$ 0.07 & 24 & 14.43 \\
HD\,202917  & 5579 $\pm$ 80  & (1)  & 4.57 $\pm$ 0.10 & 5539 $\pm$ 10 & 4.58 $\pm$ 0.06 & 0.03  $\pm$ 0.01 & This work & 5617 $\pm$ 44  & 4.39 $\pm$ 0.06 & 0.03  $\pm$ 0.03 & (3)  & 5592 $\pm$ 79  & 4.31 $\pm$ 0.17 & -0.04 $\pm$ 0.08 & 24 & 14.75 \\
WASP-3      & --             & --   & --              & 6423 $\pm$ 44 & 4.42 $\pm$ 0.15 & 0.04  $\pm$ 0.04 & This work & 6400 $\pm$ 100 & 4.25 $\pm$ 0.05 & 0.00  $\pm$ 0.20 & (9) & 6448 $\pm$ 123 & 4.49 $\pm$ 0.08 & -0.02 $\pm$ 0.08 & 25 &15.21 \\
HD\,30652   & 6499 $\pm$ 80  & (1)  & 4.33 $\pm$ 0.10 & 6494 $\pm$ 44 & 4.29 $\pm$ 0.15 & 0.04  $\pm$ 0.04 & This work & 6424 $\pm$ 44  & 4.07 $\pm$ 0.06 & 0.00  $\pm$ 0.03 & (3)  & -- & -- & -- & --  & 17.01 \\
CoRoT-3     & --	     & -- & --              & 6558 $\pm$ 44 & 4.25 $\pm$ 0.15 & 0.14  $\pm$ 0.04 & This work & 6740 $\pm$ 140 & 4.22 $\pm$ 0.07 & -0.02 $\pm$ 0.06 & (11) & -- & -- & -- & -- & 18.46 \\
XO-3        & --             & -- & --              & 6781 $\pm$ 44 & 4.23 $\pm$ 0.15 & -0.08 $\pm$ 0.04 & This work & 6429 $\pm$ 50  & 3.95 $\pm$ 0.06 & -0.20 $\pm$ 0.02 & (12) & -- & -- & -- & -- & 18.77 \\
HAT-P-41    & --             & -- & --              & 6479 $\pm$ 51 & 4.39 $\pm$ 0.22 & 0.13  $\pm$ 0.05 & This work & 6390 $\pm$ 100 & 3.68 $\pm$ 0.06 & 0.21  $\pm$ 0.10 & (13) & -- & -- & -- & -- & 20.11 \\
HAT-P-2     & --             & -- & 4.22 $\pm$ 0.10 & 6414 $\pm$ 51 & 4.18 $\pm$ 0.22 & 0.04  $\pm$ 0.05 & This work & 6290 $\pm$ 60  & 4.16 $\pm$ 0.03 & 0.14  $\pm$ 0.08 & (14) & -- & -- & -- & -- & 20.50 \\
HAT-P-34    & --             & -- & --              & 6509 $\pm$ 51 & 4.24 $\pm$ 0.22 & 0.08  $\pm$ 0.05 & This work & 6442 $\pm$ 88  & 3.98 $\pm$ 0.10 & 0.22  $\pm$ 0.04 & (15) & -- & -- & -- & -- & 24.08 \\
HD\,8673    & --             & -- & 4.29 $\pm$ 0.10 & 6472 $\pm$ 51 & 4.27 $\pm$ 0.22 & 0.14  $\pm$ 0.05 & This work & 6340 $\pm$ 44  & 4.21 $\pm$ 0.06 & 0.07  $\pm$ 0.03 & (3) & -- & -- & -- & -- & 26.91 \\
HD\,82558   & --             & -- & 4.63 $\pm$ 0.11 & 4934 $\pm$ 70 & 4.50 $\pm$ 0.21 & -0.14 $\pm$ 0.11 & This work & 5062 $\pm$ 44  & 5.12 $\pm$ 0.06 & -0.21 $\pm$ 0.03 & (3)  & -- & -- & -- & -- & 26.97 \\
CoRoT-11    & --             & -- & --              & 6343 $\pm$ 72 & 4.27 $\pm$ 0.30 & 0.04  $\pm$ 0.03 & This work & 6440 $\pm$ 120 & 4.22 $\pm$ 0.23 & -0.03 $\pm$ 0.08 & (16) & -- & -- & -- & -- & 36.72 \\
HD\,64685   & 6907 $\pm$ 80  & (1)  & 4.15 $\pm$ 0.12 & 6702 $\pm$ 98 & 3.97 $\pm$ 0.20 & -0.14 $\pm$ 0.08 & This work & 6995 & 4.36 & 0.00 & 4  & -- & -- & -- & -- & 41.59 \\
30\,Ari\,B  & 6396 $\pm$ 80  & (1)  & 4.39 $\pm$ 0.12 & 6284 $\pm$ 60 & 4.35 $\pm$ 0.25 & 0.12  $\pm$ 0.06 & This work & 6314 $\pm$ 55  & 4.29 $\pm$ 0.07 & 0.11  $\pm$ 0.04 & (17) & -- & -- & -- & -- & 42.61 \\
HD\,219877  & 6741 $\pm$ 80  & (1)  & 4.11 $\pm$ 0.13 & 6620 $\pm$ 137 & 4.10 $\pm$ 0.18 & -0.01 $\pm$ 0.06 & This work & 6775 & 4.06 & -0.13 & 4  & -- & -- & -- & -- & 54.00 \\
%---------------------------------
\hline
\end{tabular}
%}
}
%\rotatebox{90}{
\scalebox{0.95}{
\tablefoot{(1)~\citet{casa2}; (2)~\citet{eylen2014}; (3)~\citet{valenti05}; (4)~\citet{gray2006}; (5)~\citet{noyes2008}; (6)~\citet{bruntt2004}; 
(7)~\citet{bruntt2010}; (8)~\citet{huber2013}; (9)~\citet{pollacco08}; (10)~\citet{bakos2011}; (11)~\citet{deleuil2008}; 
(12)~\citet{johnskrull2008}; (13)~\citet{hartman2012}; (14)~\citet{pal2010}; (15)~\citet{bakos2012}; (16)~\citet{gandolfi2010}
(17)~\citet{prugniel2011}; (18)~\citet{sousa1}; (19)~\citet{santos2005}; (20)~\citet{ammler09}; (21)~\citet{san1}; (22)~\citet{takeda2007};
(23)~\citet{santos13}; (24)~\citet{almeida2009}; (25)~\citet{montalto2012}; }}
%}
\end{table}  
%\clearpage
\end{landscape}

\end{appendix}

\end{document}